\documentclass[lettersize,journal]{IEEEtran}
\usepackage{amsmath,amsfonts}
\usepackage{algpseudocode}
\usepackage[linesnumbered,ruled,vlined]{algorithm2e}
\usepackage{array}
\usepackage[caption=false,font=normalsize,labelfont=sf,textfont=sf]{subfig}
\usepackage{textcomp}
\usepackage{stfloats}
\usepackage{url}
\usepackage{verbatim}
\usepackage{graphicx}
\usepackage{cite}
\PassOptionsToPackage{svgnames}{xcolor}
\usepackage{xcolor}
\usepackage{tikz}

\usepackage{braket}

\usepackage{mfirstuc}

\usepackage{gensymb}
\usepackage{soul}

\usepackage{hyperref} 

\newcommand{\wymod}[1]{\textcolor{black}{#1}}
\newcommand{\modify}[1]{\textcolor{black}{#1}}
\newcommand{\secondmod}[1]{\textcolor{black}{#1}}
\begin{document}

\newcommand{\techName}[1]{\textit{PonziLens+}}
\newcommand{\Mfeature}[1]{Path Feature Module}
\newcommand{\Mgroup}[1]{Path Grouping Module}
\newcommand{\Mpath}[1]{Execution Detail Module}
\newcommand{\actW}{\includegraphics[scale=0.25]{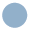}}%
\newcommand{\actP}{\includegraphics[scale=0.25]{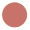}}%
\newcommand{\actC}{\includegraphics[scale=0.25]{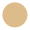}}%
\newcommand{\actR}{\includegraphics[scale=0.25]{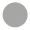}}%
\newcommand{\darkgreen}{\includegraphics[scale=0.2]{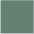}}%
\newcommand{\lightgreen}{\includegraphics[scale=0.2]{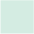}}%
\newcommand{\greyblock}{\includegraphics[scale=0.2]{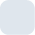}}%
\newcommand{\invest}{\includegraphics[scale=0.2]{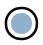}}%
\newcommand{\payment}{\includegraphics[scale=0.2]{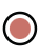}}%
\newcommand{\loopicon}{\includegraphics[scale=0.15]{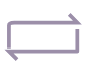}}%
\newcommand{\reward}{\includegraphics[scale=0.15]{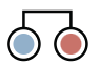}}%
\newcommand{\actwithletter}{\includegraphics[scale=0.3]{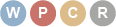}}%
\newcommand{\update}{\includegraphics[scale=0.3]{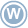}}%
\newcommand{\payback}{\includegraphics[scale=0.3]{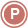}}%

\newcommand{\paymenticon}{\includegraphics[scale=0.3]{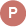}}%
\newcommand{\wicon}{\includegraphics[scale=0.3]{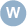}}%

\newcommand{\arrayicon}{\includegraphics[scale=0.2]{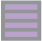}}%
\newcommand{\variable}{\includegraphics[scale=0.2]{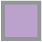}}%
\newcommand{\mapping}{\includegraphics[scale=0.2]{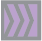}}%
\newcommand{\content}{\includegraphics[scale=0.3]{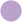}}%
\newcommand{\caller}{\includegraphics[scale=0.23]{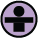}}%
\newcommand{\callvalue}{\includegraphics[scale=0.23]{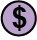}}%

\title{\techName{}: Visualizing Bytecode Actions for Smart Ponzi Scheme Identification}

\author{Xiaolin Wen \orcid{0000-0002-8562-7640}, 
Tai D. Nguyen\orcid{0000-0003-2109-845X}, 
Shaolun Ruan\orcid{0000-0002-6163-9786}, 
Qiaomu Shen\orcid{0000-0002-6510-0964}, 
Jun Sun\orcid{0000-0002-3545-1392}, 
Feida Zhu\orcid{0000-0001-6077-4356},
and Yong Wang\orcid{0000-0002-0092-0793}
\thanks{X. Wen and Y. Wang are with Nanyang Technological University. Email: xiaolin004@e.ntu.edu.sg, yong-wang@ntu.edu.sg. Part of this work was done when they were affiliated with Singapore Management University.} 
\thanks{T. Nguyen, S. Ruan, J. Sun, and F. Zhu are with Singapore Management University. Email: \{dtnguyen.2019, junsun, fdzhu\}@smu.edu.sg, slruan.2021@phdcs.smu.edu.sg.}
\thanks{Q. Shen is with Southern University of Science and Technology. Email: shenqm@sustech.edu.cn.
}
\thanks{Y. Wang and Q. Shen are the corresponding authors.}
}
\markboth{Journal of \LaTeX\ Class Files,~Vol.~14, No.~8, August~2021}%
{Shell \MakeLowercase{\textit{et al.}}: A Sample Article Using IEEEtran.cls for IEEE Journals}


 \maketitle

\begin{abstract}
With the prevalence of smart contracts, smart Ponzi schemes have become a common fraud on blockchain and have caused significant financial loss to cryptocurrency investors in the past few years.
Despite the critical importance of detecting smart Ponzi schemes,
a reliable and transparent identification approach adaptive to various smart Ponzi schemes is still missing.
\modify{To fill the research gap, we first extract semantic-meaningful actions to represent the execution behaviors specified in smart contract bytecodes, which are derived from a literature review and in-depth interviews with domain experts. 
We then propose \techName{}, a novel visual analytic approach that provides an intuitive and reliable analysis of Ponzi-scheme-related features within these execution behaviors.
\techName{} has three visualization modules that intuitively reveal all potential behaviors of a smart contract, highlighting fraudulent features across three levels of detail.
It can help smart contract investors and auditors achieve confident identification of any smart Ponzi schemes.}
We conducted two case studies and in-depth user interviews with 12 domain experts and common investors to evaluate \techName{}.
The results demonstrate the effectiveness and usability of \techName{} in \modify{achieving} an effective identification of smart Ponzi schemes.
\end{abstract}


\section{Introduction}







\IEEEPARstart{P}{onzi} scheme~\cite{artzrouni2009mathematics} is a classic financial fraud that appeared in the offline world over 150 years ago. It uses the investment of new investors to compensate the existing investors, and will inevitably collapse and cause financial losses to most investors.
With the rapid development of blockchain technology, Ponzi schemes have also become widely spread on blockchain platforms (e.g., Ethereum\footnote{\url{https://ethereum.org/}}) in the past few years~\cite{bartoletti2020dissecting,mukherjee2021cryptocurrency}.
Such Ponzi schemes leverage smart contracts on the blockchain and are run in a decentralized, anonymous, and immutable manner, which are called \textbf{smart Ponzi schemes}~\cite{chen2018detecting}.
\modify{
Smart Ponzi schemes often lure investors by posing as high-profit investment plans~\cite{chen2018detecting}. 
For example, Forsage, a platform marketed as a low-risk investment using smart contracts, was revealed to be a \$340 million global Ponzi scheme~\cite{Forsage}.
At present, smart Ponzi schemes have become one of the most common frauds for cryptocurrency~\cite{fan2021spsd, FENG2024108868}.
Ponzitracker~\cite{Ponzitracker} reported that the total investor funds at risk in smart Ponzi schemes reached nearly \$3 billion in 2022.
Such significant financial losses to investors seriously affect the development of the blockchain ecosystem~\cite{liang2024ponzicuard, zheng2023securing}. 
}

\begin{figure}[!htbp]
\centering
\includegraphics[width=3.45in]{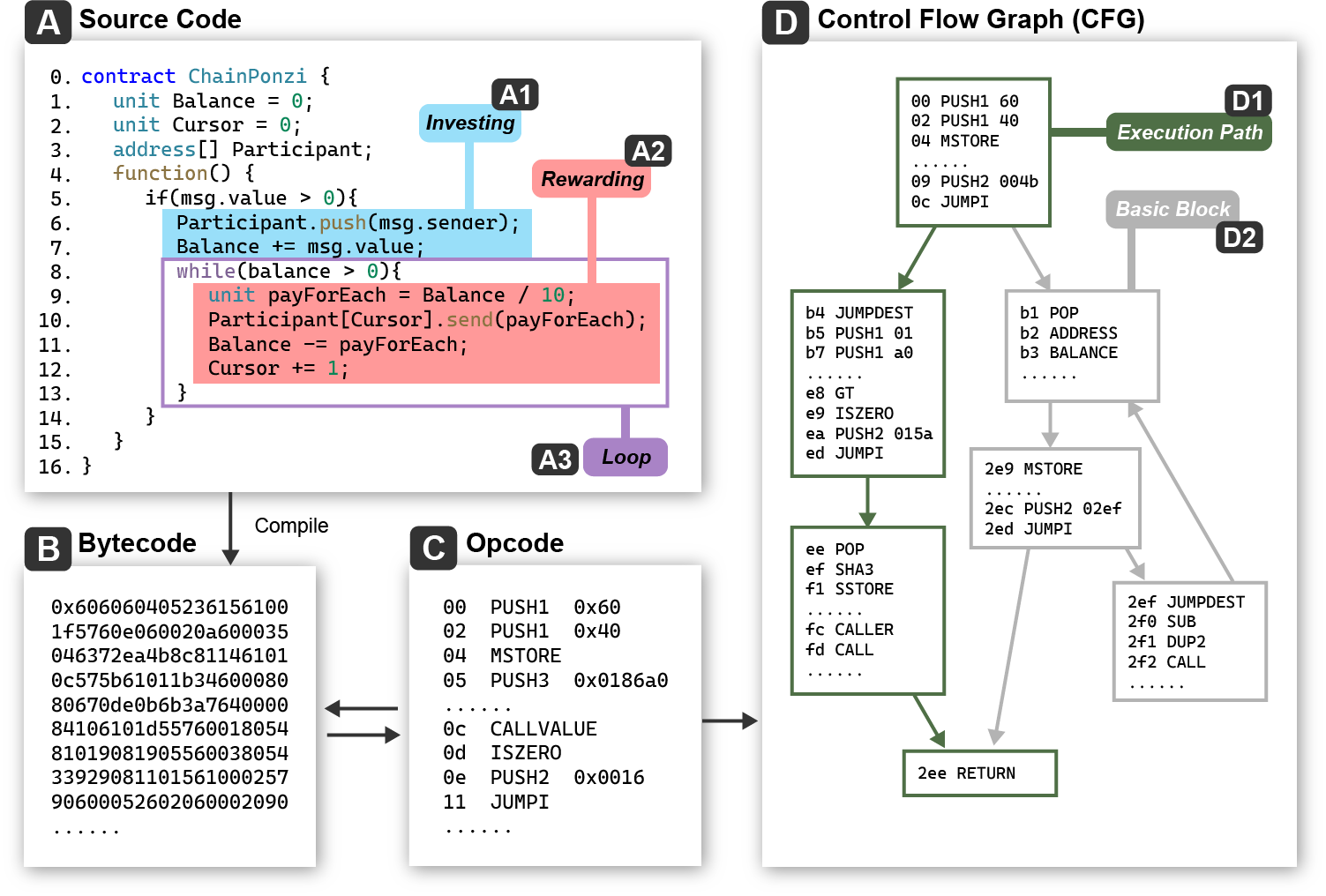}
\caption{The showcase of critical concepts in this study: (A) shows an example source code of a chain-type smart Ponzi scheme with investing behavior (A1) and rewarding (A2) behavior in a loop (A3). (B) and (C) show the bytecode and opcode of a smart contract. (D) demonstrates the control flow graph with execution paths (D1) and basic blocks (D2).}
\label{background}
\end{figure}

\modify{
Although smart contract codes are publicly available on the blockchain, investors often struggle to verify them and end up trusting them blindly due to their lack of ability to thoroughly inspect the code~\cite{bartoletti2020dissecting}.
To protect investors, various methods have been proposed to automatically detect smart Ponzi schemes~\cite{chen2019exploiting,wang2021ponzi,liang2021data,galletta2023sharpening,chen2018detecting,fan2021spsd,chen2021sadponzi,liang2024ponzicuard,lu2024Soucep}.}
Early research focused on analyzing the transaction records on the blockchain and extracting features related to Ponzi schemes, such as transaction frequency, amount, and networks~\cite{chen2019exploiting,wang2021ponzi,liang2021data,galletta2023sharpening}. 
However, these approaches intrinsically depend on the transaction data, meaning that some investors have already finished the immutable transactions \modify{and suffered from financial losses.
To achieve early detection before transactions occur, some approaches have focused on detecting Ponzi schemes using the smart contract code (Fig.~\ref{background}A-C), including the source code, opcode, and bytecode}~\cite{chen2018detecting,bartoletti2020dissecting,fan2021spsd,chen2021sadponzi,wen2023code,liang2024ponzicuard}. 
The above code-based 
approaches for smart Ponzi scheme detection still suffer from two major challenges:
\modify{
\textbf{(C1) limited adaptability:} 
These approaches inherently rely on fixed rules or models trained on existing labeled smart contracts. 
However, scammers continuously create new Ponzi scheme variants that evade these rules or criteria~\cite{zheng2023securing}\modify{, making these ad-hoc} approaches ineffective against new smart Ponzi schemes~\cite{chen2021sadponzi}.
\textbf{(C2) lack of transparency:} 
These approaches leverage machine learning models, combined with abstract features of transaction records or codes, to classify smart contracts as smart Ponzi schemes without providing understandable and reliable evidence~\cite{FENG2024108868}. 
This makes it difficult for common investors to make confident investment decisions. 
Our preliminary work, \textit{PonziLens}\cite{wen2023code},
attempted to visually explain Ponzi features at the opcode execution level, but it is still difficult for common investors to understand the complex opcode of smart contracts and a more intuitive approach is still missing~\cite{li2020stan}.
}

\modify{To address these challenges, we propose \textbf{\techName{}}, a novel visual analytic approach to help smart contract investors and auditors identify smart Ponzi schemes intuitively.
\techName{} can reveal the semantic meaning of the smart contract execution process and allow users to conveniently validate various smart contracts across three levels: contract, execution path group, and execution path.
Specifically, we first conduct a preliminary study with four domain experts to derive design requirements and propose a framework for semantic action extraction.
This framework translates complex smart contract bytecodes into semantic action sequences, i.e., a series of actions with meaningful semantics, to indicate the possible behaviors of a smart contract.
These action sequences allow users to identify fraudulent behaviors and further verify whether a smart contract is a Ponzi scheme, instead of fully relying on fixed rules (\textbf{C1}).
Furthermore, we propose a novel hierarchical visualization design including three modules
to show the action sequences of a smart contract in a top-down manner, intuitively revealing the smart contract behaviors as well as the Ponzi scheme-related features (\textbf{C2}).}
With the help of \techName{},
users can gain deep insights into the detailed behaviors of any smart contracts and identify smart Ponzi schemes \modify{intuitively and reliably.}
%
To evaluate the effectiveness and usability of \techName{}, we conducted two case studies and in-depth interviews with domain experts and common investors.
\modify{The results show that \techName{} is effective for intuitively identifying smart Ponzi schemes.
Additionally, \techName{} also has potential applications in source code understanding~\cite{pierro2021smart} and software testing~\cite{zhou2019visfuzz} beyond Ponzi schemes. 
To the best of our knowledge, we are the first to visualize smart contract codes for identifying smart Ponzi schemes.}
Our major contributions are listed as follows:
\begin{itemize} 
    \item  We formulate the design requirements through collaboration with domain experts and \modify{propose a framework to extract semantic-meaningful actions} to delineate the action sequences
    during the execution of smart contracts. 
    \item We design \techName{}, a novel visual analytic approach with three visualization modules to 
    facilitate a comprehensive top-down analysis of Ponzi-scheme-related features in smart contracts,
    enabling adaptive and reliable identification of smart Ponzi schemes.
    %
    
    \item We present two case studies and conduct in-depth user interviews with both domain experts and common investors to illustrate the effectiveness and usability of \techName{}.
\end{itemize}


\section{Related Work}
\modify{
This work is related to prior research on
\textit{smart Ponzi scheme detection}, \textit{visual identification of blockchain frauds}, and \textit{software visualization}}.

\subsection{Smart Ponzi Scheme Detection}
Early studies on smart Ponzi scheme detection relied on manual checks of smart contracts' source codes~\cite{bartoletti2020dissecting}. 
Recently, automatic smart Ponzi scheme detection algorithms were proposed, which can mainly be divided into \textit{transaction-based} and \textit{code-based}, according to their inputs~\cite{chen2021sadponzi}. 

\textbf{\textit{Transaction-based}} methods detect Ponzi schemes by extracting information from the existing transaction records of smart contracts. 
For instance, prior studies~\cite{chen2019exploiting,wang2021ponzi, galletta2023sharpening} have defined different features of transactions (e.g., payment frequency and balance) and further trained customized models with these features to
automatically detect Ponzi schemes from the given smart contracts.
\modify{Besides, Liang et al.~\cite{liang2021data} extracted the structure features of transaction networks and incorporated them in their Ponzi scheme detection model.}
These approaches intrinsically rely on transaction records, making them unable to achieve the early detection of Ponzi schemes before investors are really trapped in smart Ponzi schemes.

\modify{\textbf{\textit{Code-based}} methods aim to detect smart Ponzi schemes before executing transactions by analyzing the code information (e.g., source code, opcode, and bytecode) available once the smart contract is deployed.
Initially, these methods focused on relatively simple features like opcode frequency or source code text~\cite{bartoletti2020dissecting,lou2020ponzi,ibba2021evaluating, fan2020mask,fan2021spsd}.}
\modify{However, reducing intricate codes to simple features results in significant information loss, making it difficult to fully capture runtime information and rendering these methods unreliable for Ponzi scheme detection.}
Accordingly, recent studies have utilized \modify{ \textit{Control Flow Graphs (CFG)}, which graphically represent a smart contract's execution paths} to capture its execution logic~\cite{sun2020early,chen2021sadponzi,liang2024ponzicuard}.
However, these approaches still lack an understandable explanation for why a smart contract is classified as a Ponzi scheme. 
\modify{Also, these methods are based on fixed patterns, rules, or models trained on existing labeled smart contracts, making it hard to detect new variants of Ponzi schemes.}
\techName{} can visualize the bytecode execution process of a smart contract as semantically meaningful action sequences with highlighted Ponzi-scheme-related features.
This allows users to deeply understand why a smart contract is a Ponzi scheme or not and identify new variants by checking the detailed behaviors.

\subsection{Visual Identification of Blockchain Frauds}
\modify{Due to the lack of effective regulation 
and complex data in blockchain application scenarios,
numerous visualizations have been developed to help people understand blockchain data and detect fraud visually~\cite{tovanich2019visualization}.}

\modify{Most studies focus on fraud within blockchain transactions, identifying issues like coin mixing~\cite{di2015bitconeview}, stealing~\cite{ahmed2019tendrils}, and money laundering~\cite{bistarelli2017go} by visualizing value flow and tracking cryptocurrency movement across entities over time.}
Other frauds can be recognized by visualizing features in the transaction networks, such as high-frequency transactions~\cite{mcginn2016visualizing} and wash trading~\cite{wen2023nftdisk}.
Also, visualizing the networks between various entities (e.g., addresses~\cite{isenberg2017exploring}, clusters~\cite{kinkeldey2017bitconduite}, and exchanges~\cite{yue2018bitextract}) in the blockchain
can help identify some abnormal structures and entities to avoid potential fraud.

In addition to frauds occurring in transaction records, 
there are also frauds deployed on the blockchain through smart contracts.
These frauds can be detected before transactions occur by analyzing the code of smart contracts, including Ponzi schemes (the focus of this study). 
Before our study, there were no visualization tools designed for detecting fraud within smart contract codes.
While there are existing visualizations aimed at aiding in understanding control flow graphs~\cite{norvill2018visual,devkota2018cfgexplorer,devkota2021cfgconf}, none of them focus on fraud detection.
Our preliminary study~\cite{wen2023code} aimed to visualize the execution paths in the original control flow graph and revealed the investment and rewarding flows by recognizing critical opcodes. 
Nevertheless, it is still difficult for common investors to understand opcodes and conduct an in-depth analysis of smart contract codes. 
Therefore, in this paper, we move one step further by extracting semantically-meaningful actions from the opcodes of smart contracts, which allows common investors to gain deep insights into smart contracts and easily identify smart Ponzi schemes.



\subsection{Software Visualization}
Smart contracts are software on the blockchain~\cite{zhou2023security}, which makes this work also relevant to software visualization.
According to the visualized features of software systems, previous software visualizations fall into three categories~\cite{diehl2007software,chotisarn2020systematic}: structure, evolution, and behavior. 
Structure visualizations aim to analyze
information like
source code~\cite{hayatpur2023crosscode}, package dependencies~\cite{isaacs2018preserving}, and control flow graphs~\cite{devkota2018cfgexplorer,devkota2021cfgconf}, to aid in understanding software structure.
Evolution visualizations analyze the code change history to show the evolution of a software system~\cite{yoon2013visualization,kim2020githru}.
Behavior visualizations display data collected from program execution like function calls~\cite{zhou2024fctree} to facilitate performance optimization~\cite{isaacs2014combing} and anomaly detection~\cite{xu2019clouddet}.
Existing behavior visualizations typically display time series~\cite{sun2021daisen} or domain-specific event sequences~\cite{isaacs2014combing,guo2018valse}.
Our work belongs to the behavior visualization of software, as smart contract behaviors during execution are visualized.
\secondmod{
Unlike existing \wymod{software} behavior visualization methods, our approach introduces a novel hierarchical visual design that represents potential smart contract behaviors across three levels of \wymod{granularity}. This design emphasizes Ponzi scheme features at different levels of \wymod{details}, enabling users to efficiently audit a smart contract without navigating complex source code.
}

\section{Background}
This section introduces the background, including smart contracts, control flow graphs, and smart Ponzi schemes.
\subsection{Smart Contracts}
\textbf{\textit{Smart contracts} }are self-executing programs deployed on a blockchain platform, which will be automatically executed when specific predefined trigger events are met.
\secondmod{
Smart contracts have been widely utilized across various domains, including automated payments and asset exchanges on decentralized finance (DeFi) platforms ~\cite{el2023smart}.
}
\textit{Ethereum}~\cite{wood2014ethereum} is one of the most well-known blockchain platforms with smart contracts incorporated.
\secondmod{While other blockchains also support smart contracts}, we focus on Ethereum in this study.

Smart contracts can be written in various programming languages like Solidity, Viper, and Serpent~\cite{soud2023dissecting}.
The code written in these languages is called the \textbf{\textit{source code}} of smart contracts, and Fig.~\ref{background}A shows an example of the Solidity source code of a smart contract.
When deploying these smart contracts on the blockchain, they are compiled into the hexadecimal \textbf{\textit{bytecode}} (Fig.~\ref{background}B) and sent to the blockchain.
\secondmod{
Bytecode can be converted into a sequence of instructions written in \textbf{\textit{opcode}}~\cite{Opcodes} (i.e., ``\textit{operation code}") and operands, as shown in Fig.~\ref{background}C. 
For instance, the bytecode ``\textit{Ox6000}" is equal to an instruction with opcode ``\textit{PUSH1}" and operand ``\textit{0x00}".
The bytecodes of all smart contracts are accessible on the blockchain, but not all source codes are publicly released.
 This is why we chose bytecodes as the input for our study.}

\subsection{Control Flow Graphs (CFG)}
\secondmod{
In Ethereum, smart contracts execute within the \textit{\textbf{Ethereum Virtual Machine (EVM)}}, a stack-based run-time environment. 
Similar to traditional programs, smart contracts use the \textit{\textbf{stack}} for temporary data (e.g., holding variables during function calls), employ \textit{\textbf{memory}} for short-term data storage within a transaction (similar to RAM,  Random-Access Memory), and utilize \textit{\textbf{storage}} for persistent data that remains accessible across transactions (like global variables)~\cite{wood2014ethereum}.
Most opcodes take the values at the stack top as operands and place the results back onto the stack. Among them, several opcodes can modify the memory (e.g., \textit{MSTORE}) and storage (e.g., \textit{SLOAD} and \textit{SSTORE})~\cite{Opcodes}.
}

\secondmod{
During actual execution in EVM, smart contracts execute different code paths based on conditional statements (e.g., \textit{if} and \textit{else}), loops, and function calls.
\textbf{\textit{CFG}} (Fig.~\ref{background}D) is a static graphical representation of all potential execution paths (control flows) of a smart contract~\cite{contro2021ethersolve}.
The smallest execution unit in CFG is the \textit{\textbf{basic block}} (Fig.~\ref{background}D2), which consists of a sequence of instructions. 
Each basic block has no loops, and the directed links between basic blocks indicate possible execution orders.
When invoking a transaction, the contract executes along a sequence of basic blocks in the CFG, which is determined by the conditions.
This sequence is called the \textit{\textbf{Execution Path}} (Fig.~\ref{background}D1).
}

\subsection{Smart Ponzi Schemes}
A \textit{Ponzi scheme} is a classic financial scam in which investors are lured with promises of exceptionally high returns on their investments~\cite{artzrouni2009mathematics}. The scheme operates by using the investments from new investors to pay returns to earlier investors.
Such a game continues until there are insufficient funds left, ultimately leading to the collapse~\cite{moore2012postmodern}.
A \textit{smart Ponzi scheme} refers to a Ponzi scheme deployed on a smart contract~\cite{bartoletti2020dissecting,chen2018detecting}.
According to existing studies~\cite{bartoletti2020dissecting,chen2021sadponzi,liang2024ponzicuard}, the \textbf{features} of smart Ponzi schemes are as follows: 
\begin{itemize}
    \item \textbf{F1. Include \textit{investing} and \textit{rewarding} behaviors.} Investing indicates that after receiving an investment the contract records investment information like investor's addresses and investing amount. Rewarding means that the contract redistributes money among previous investors.
    \item \textbf{F2. Receive money only from investors.} This feature can rule out smart contracts with external asset sources, like a bank that pays the interest of a ``smart" bond or enterprises that distribute incentives to employees.
    \item \textbf{F3. Guarantee higher profits for all participants.} Each investor can make a profit if there is enough money in the contract afterward. This rules out contracts like gambling or games, where not all investors can get high rewards.
\end{itemize}

Fig.~\ref{background}A shows a smart Ponzi scheme written in Solidity. 
For F1, Lines 6-7 (Fig.~\ref{background}A1) indicate that the contract stores the investor's address (\textit{msg.sender}) into an array called \textit{Participant} and updates the \textit{Balance} with the investment amount (\textit{msg.value}) under the condition that the investment amount is greater than zero (Line 5), \modify{which is an investing behavior. 
In Fig.~\ref{background}A2, the contract uses a loop (A2) to pay 10\% of the \textit{Balance} to each previous investor, which is a rewarding behavior.}
This contract aligns with F2 as it contains only one function for receiving investments and distributing rewards to previous investors.
If there are enough new investments, all the previous investors can get high rewards, satisfying F3.
Therefore, this contract is indeed a smart Ponzi scheme.



\section{Informing the Design}
To achieve a rational design, we conducted a preliminary study to gather the requirements for guidance.
This section reports the details and feedback of our preliminary study.

\subsection{Preliminary Study}
During the preliminary study, we interviewed four domain experts from both academia and industry, with rich experience in smart Ponzi scheme detection.
E1 and E2 are university professors whose research areas encompass web3 fraud detection and smart contract analysis.
E3 works as a smart contract auditor at a web3 security company, while E4 serves as a business analyst in another web3 company with services in smart contract auditing.
All four experts have strong backgrounds in smart contracts and diverse insights into smart Ponzi schemes.
We conducted open-ended interviews with each expert individually, with each session lasting about one hour.
The interview was constructed into three phases, each focusing on a specific aspect:
\textit{Necessity of Our Study}, \textit{Ponzi Features at Bytecode Level}, and \textit{Design Requirements}.
The feedback about the necessity is introduced in the introduction and related work sections, while the Ponzi features and design requirements are described in the subsequent sections.

\subsection{Ponzi Features at Bytecode Level}\label{sec:background}
According to the previous studies~\cite{bartoletti2020dissecting,chen2021sadponzi}, smart Ponzi schemes can be roughly divided into four types: 
\modify{
\begin{itemize}
    \item \textit{\textbf{Handover-scheme}}: Upon receiving investments from a new investor, the contract will proceed to send profits to the previous investor (the last one who has invested) and subsequently update the address of the new investor as the new ``last investor" in the scheme.
    \item \textit{\textbf{Chain-scheme}}: Upon receiving investments from a new investor, the contract will distribute profits to all the previous investors stored in a chain-like structure.
    Then, the contract will add the new investor to this chain.
    \item \textit{\textbf{Tree-scheme}}: In a tree-scheme, investors are organized in a tree-like structure based on the invitation relationships. Upon receiving an investment, the contract will sequentially distribute profits to the previous inviters of each node in the tree. 
    \item \textit{\textbf{Withdraw-scheme}}: In a withdraw-scheme, the amount of each investor's balance is stored in the blockchain. When investments are received, it will increase the balance of each investor, instead of the direct redistribution. Investors have to invoke another execution to withdraw their money according to their balance. 
\end{itemize}
}
During interviews, domain experts pointed out that F1 plays a pivotal role in identifying Ponzi schemes in terms of contract bytecodes, i.e., the identification of investing and rewarding patterns, which helps eliminate non-Ponzi (not a Ponzi scheme) contracts quickly and distinguish the types of Ponzi schemes. 
\modify{Since F2 requires analyzing investing behaviors and F3 involves examining the amount of investing and rewarding, identifying these behaviors (F1) is fundamental to both F2 and F3.}
\modify{To facilitate the identification of F1, we define four critical features at the bytecode level based on the investing and rewarding behaviors of four typical smart Ponzi schemes, which are referred to as \textit{\textbf{Ponzi Features (PF)}} in this paper.}
\begin{itemize}
    \item \textit{\textbf{(PF1) Investing}}: The execution path stores the investor's address in the EVM's storage for future reward payments.
    \item \textit{\textbf{(PF2) Payment}}: The execution path distributes funds to an entity. This Ponzi feature is helpful for the verification of fund destinations in smart contracts.
    \item \textit{\textbf{(PF3) Loop}}:
    \modify{A loop exists in the execution path. In chain-schemes and tree-schemes, loops iteratively process payments to previous investors, while in withdraw-schemes, loops update the balances of previous investors.}
    \item \textit{\textbf{(PF4) Rewarding}}: The execution path directly redistributes investment to previous investors, demonstrated through payments, with the receiver retrieved from the same slots storing the investor addresses. 
\end{itemize}

\begin{figure*}[t]
  \centering 
    \setlength{\abovecaptionskip}{-0.1cm}
  \includegraphics[width=0.85\linewidth
  ]{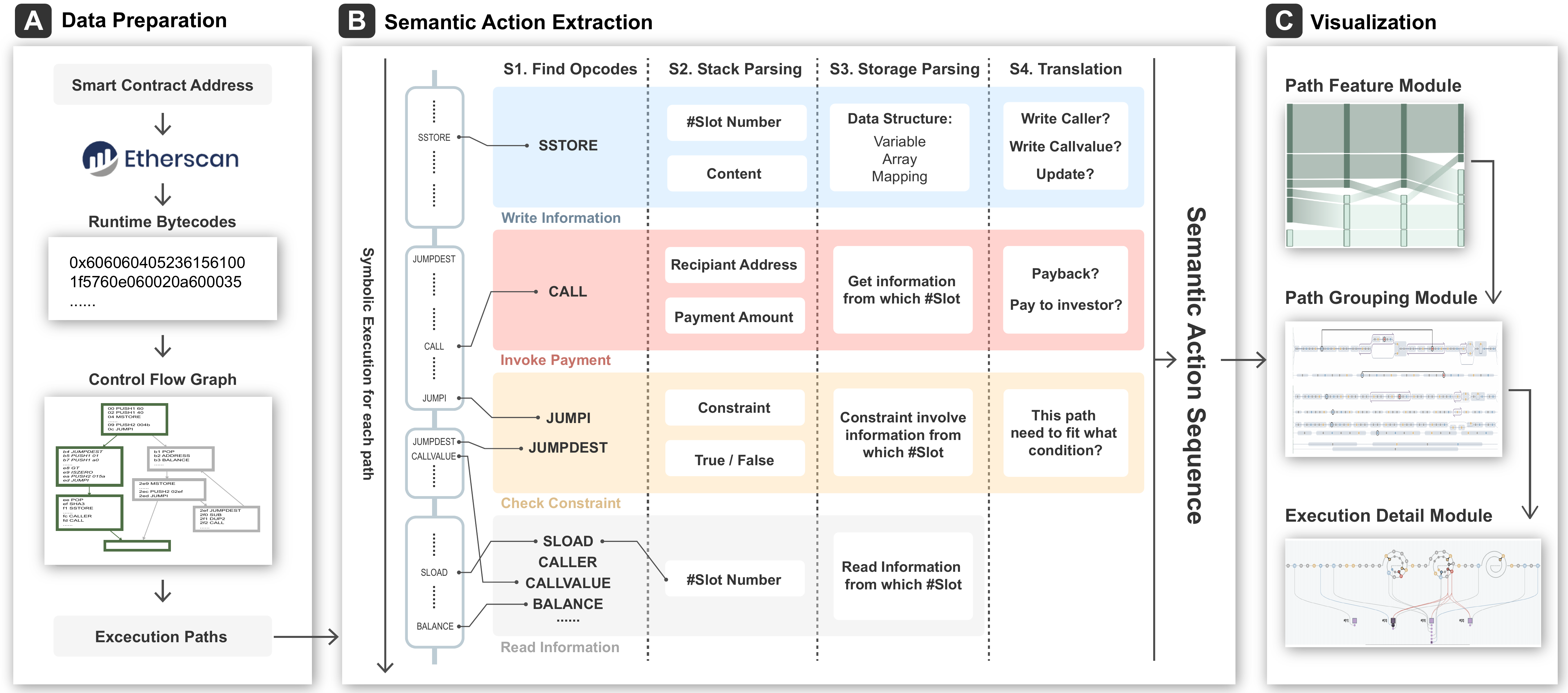}
  \caption{The framework of \techName{}. (A) shows the data preparation for the collection of potential execution paths. (B) shows the \modify{semantic action extraction that generates} semantic action sequences from each execution path. (C) demonstrates three visualization modules in \techName{}.}
  \label{workflow}
\end{figure*}

\subsection{Design Requirements}
During the preliminary study, all the experts (E1-E4) agreed that the visual identification of Ponzi schemes should follow a top-down workflow, beginning with a comprehensive understanding of the smart contract's functionality and eventually finding out the specific execution paths for conducting Ponzi schemes.
Distilling experts' feedback, we organize the design requirements into three levels: \textbf{Contract}, \textbf{Group}, and \textbf{Path}. 
The contract level \modify{provides an} overview of Ponzi features across all potential paths in a smart contract. 
The group level emphasizes action patterns within groups of execution paths of similar Ponzi features, while the path level concentrates on detailed execution information for each execution path. 

\begin{itemize}
    \item{\textbf{R1}} (Contract) \textbf{Overview the Ponzi feature distribution.} Experts (E1-E4) have noted that initiating the analysis by gaining an overview of the Ponzi feature distribution can aid in rapidly assessing the possibility of a Ponzi scheme.
    Grouping paths with similar Ponzi features helps locate the paths responsible for conducting the Ponzi scheme.   
    \item \textbf{R2} (Group) \textbf{Summarize action patterns within each execution path group.} All experts (E1-E4) mentioned that checking the action patterns of each path group can help quickly identify the functionality of the smart contract. E3 emphasized summarizing both unique and common actions among multiple execution paths can help understand the overall action patterns of each group more effectively than presenting each path individually.
    \item  \textbf{R3} (Group) \textbf{Highlight the Ponzi features of each path group.} 
    Given R2, two experts (E2, E4) commented that the visual summary should highlight Ponzi features upon the concrete actions, which can provide insight into the evidence of Ponzi schemes and help filter out critical execution paths for further analysis. 
    \item  \textbf{R4} (Path) \textbf{Show the detailed action sequences of individual paths during execution.} All experts (E1-E4) agreed that identifying a smart Ponzi scheme usually depends on the features in several paths, and it is necessary to analyze the actions of these paths during execution in detail. E3 pointed out that the time and conditions under which a path invokes a payment are crucial evidence for Ponzi scheme detection. Therefore, our method should visualize the whole action sequences of an execution path.
    \item  \textbf{R5} (Path) \textbf{Support the deep analysis of action patterns of the loop. } All experts (E1-E4) emphasized the loop is a crucial Ponzi feature appearing in three out of the four Ponzi scheme types. Our method should facilitate the comprehension of behavior within each round of loops, including whether each round within the loop performs the same or different actions. Additionally, E2 added that actions within the loop can also help identify Ponzi scheme types and understand the money flow.
    \item  \textbf{R6} (Path) \textbf{Display the storage interactions of individual paths during execution.}
    The storage stores permanent data of smart contracts, including investment information (i.e., previous investors' addresses and investment amounts) and rewarding information (i.e., recipient addresses and profit amounts). All experts (E1-E4) said that the storage interactions during execution reveal vital information for Ponzi scheme detection, such as which slots offer recipient addresses during payments.
    E4 added that the storage structure of investors' addresses indicates the Ponzi scheme types. For instance, a tree-like structure often indicates a tree-scheme, while the handover-scheme typically stores investors' addresses in variables~\cite{chen2021sadponzi}.  
    
\end{itemize}

\section{Semantic Action \modify{Extraction}}
\modify{This section introduces 
the semantic action extraction component of our approach (Fig.~\ref{workflow}B).
}

\subsection{Action Definitions}
As shown in Fig.~\ref{background}, the bytecodes of smart contracts are abstract and complex for common investors to understand.
Therefore, we define a set of critical and semantic-meaningful actions related to Ponzi schemes and transform the bytecodes of smart contracts into meaningful action sequences.
The definitions of actions ensure that the aforementioned Ponzi features can be reflected and the design requirements can be supported by these actions.
Specifically, we define four critical actions related to smart Ponzi schemes.
The actions and corresponding definitions 
are as follows:

\textbf{\textit{Write Information}} captures the action of storing information in EVM storage slots.
This action can help capture PF1, i.e., storing investment information, as well as inferring the contents and structure of storage slots (R6).

\textbf{\textit{Invoke Payment}} signifies that a smart contract issues a specific payment to a recipient, a common feature (PF2) in all the Ponzi schemes for distributing profits to investors. Tracking money flow is crucial for identifying such schemes.

\textbf{\textit{Check Constraint}} \modify{represents the specific conditions that the execution path should meet, particularly at decision points in CFGs.
For example, the path involving the investing function requires that the amount invested must exceed a specific value.}

\textbf{\textit{Read Information}} involves loading values into EVM stacks for processing, including data from storage slots or EVM context (like contract balances or current timestamps).
While this action is not tied to a specific Ponzi scheme, it aids in deducing a complete action sequence, particularly when other actions are too complex to understand.

\subsection{Data Preparation}
To identify and generate the above four actions, we first collect all feasible execution paths of a smart contract as depicted in Fig.~\ref{workflow}A. 
Given a smart contract address that users want to analyze, we start by retrieving the bytecode using EtherScan\footnote{https://etherscan.io/}, a well-known blockchain explorer. 
Subsequently, we build a CFG from these bytecodes with the aid of Teether\modify{, a smart contract testing tool~\cite{krupp2018teether}}, and traverse all paths in the graph.
According to R5, when a loop is present in a path, we collect only the paths that encompass two rounds of the loop, which reduces the repeated paths and ensures sufficient information is gathered to deduce the loop's functionality.
\modify{
Finally, we use Teether~\cite{krupp2018teether} to run symbolic execution and test each path, collecting all potential execution paths that can successfully run on the real blockchain.}

\modify{
Symbolic execution is a software testing technique that replaces normal inputs (e.g., numbers) with symbolic values (e.g., formulae) during execution~\cite{he2019learning}.
For smart contracts, actual values are replaced by symbolic ones, such as \textit{CALLER} for the investor's address and \textit{CALLVALUE} for the investment amount.
During the symbolic execution, the contents in the stacks are represented as Z3 constraints, i.e., formulas with semantic meanings that can be parsed by Z3~\cite{de2008z3}, a constraint solver. 
For example, if the top two stack contents are \textit{CALLVALUE} and \textit{1}, executing the \textit{ADD} opcode will result in a Z3 constraint \textit{CALLVALUE+1}.
By analyzing these Z3 constraints in stacks when specific opcodes are encountered, we can derive the semantic meaning of the opcode operations and construct the corresponding semantic actions.
}

\subsection{Action Sequence Generation}\label{Sec:action}
We select a subset of opcodes related to the four actions and construct a semantic action sequence for each collected path.
Specifically, we extract actions by analyzing the content in the EVM when encountering specific opcodes during the above symbolic execution, following four steps (S1-S4) shown in Fig.~\ref{workflow}B.
Steps S1-S3 follow the bytecode execution process on the EVM, where each opcode first manipulates the stack and then interacts with storage, while S4 aims to make the generated actions understandable for users.

\textbf{S1. Find Specific Opcodes:} Each action is associated with specific opcodes. For example, \textit{SSTORE} is used for \textit{\wicon Writing Information}, while \textit{CALL} is used to \textit{\paymenticon Invoke Payment}.
Upon encountering any of these opcodes during symbolic execution, we \modify{ can initially identify the four action types \actwithletter \, and then conduct S2-S4 to enrich its semantic information.
}

\textbf{S2. Stack Parsing:} \modify{In the EVM stack, we extract the current operation parameters for the above opcodes.
For example, the two parameters of \textit{SSTORE} indicate what content \content \, is stored in which storage slot for \textit{\wicon Write Information}.
With \textit{CALL}, we identify the recipient's address and the payment amount for \textit{\paymenticon Invoke Payment}.
\secondmod{
These parameters are represented as symbols in the format of Z3 constraints~\cite{de2008z3} during symbolic execution, instead of concrete numerical values, and are interactively shown in the tooltip in \techName{}.
}}

\textbf{S3. Storage Parsing:} By analyzing these Z3 constraints \modify{collected in S2, we can collect the relationships between storage slots and actions, i.e., where these values are collected from or related to which slots, and the slot structures.
\secondmod{(The relationships between storage slots and actions are represented as the links between actions and storage slots in the \Mpath{})}.} For instance, \textit{\paymenticon Invoke Payment} retrieves the recipient's address from which slot or calculates the payment amount depending on values from which slot. For \textit{\wicon Write Information}, we can recognize the data structures (e.g., variable \variable \, or array \arrayicon) of storage slots by the different Z3 constraints for slot numbers.

\textbf{S4. Translation:} 
\modify{With S1-S3, we can depict actions from the perspective of program execution, such as writing specific content into a particular slot.
S4 aims to translate the information collected from S1-S3, especially Z3 constraints, into understandable semantics, further aiding users in identifying Ponzi scheme features.
For example, if the content written by \textit{Write Information} is identified as the investor’s address, we mark it as \textit{\invest Investing (PF1)} and record the target slot storing investor addresses. Its slot structure can help identify the chain or tree types of Ponzi schemes. If an \textit{\payment Invoke Payment} action collects the payee's address from this slot, it is marked as \textit{\reward Rewarding (PF4)}. 
Additionally, if \textit{Write Information} writes content that includes the previous content of the same slot, we mark it as ``\textit{\update Update Information}", commonly used in Ponzi schemes for updating investor counts or balance.
If \textit{Invoke Payment}'s payee is the current investor \textit{CALLER \caller}, we label it as ``\textit{\payback Payback}", which often occurs when investors withdraw money, indicating a withdraw-scheme, or return their investment when certain conditions are not met.
}

\modify{
Through these four steps, we collected action types (S1), operands (S2), storage slot interactions (S3), and relations with Ponzi features and additional semantics (S4), representing each execution path as a sequence of semantic actions.
}


\begin{figure*}[!htbp]
  \centering 
    \setlength{\abovecaptionskip}{-0.2cm}
  \setlength{\belowcaptionskip}{-4cm}
  \includegraphics[width=0.9\linewidth
  ]{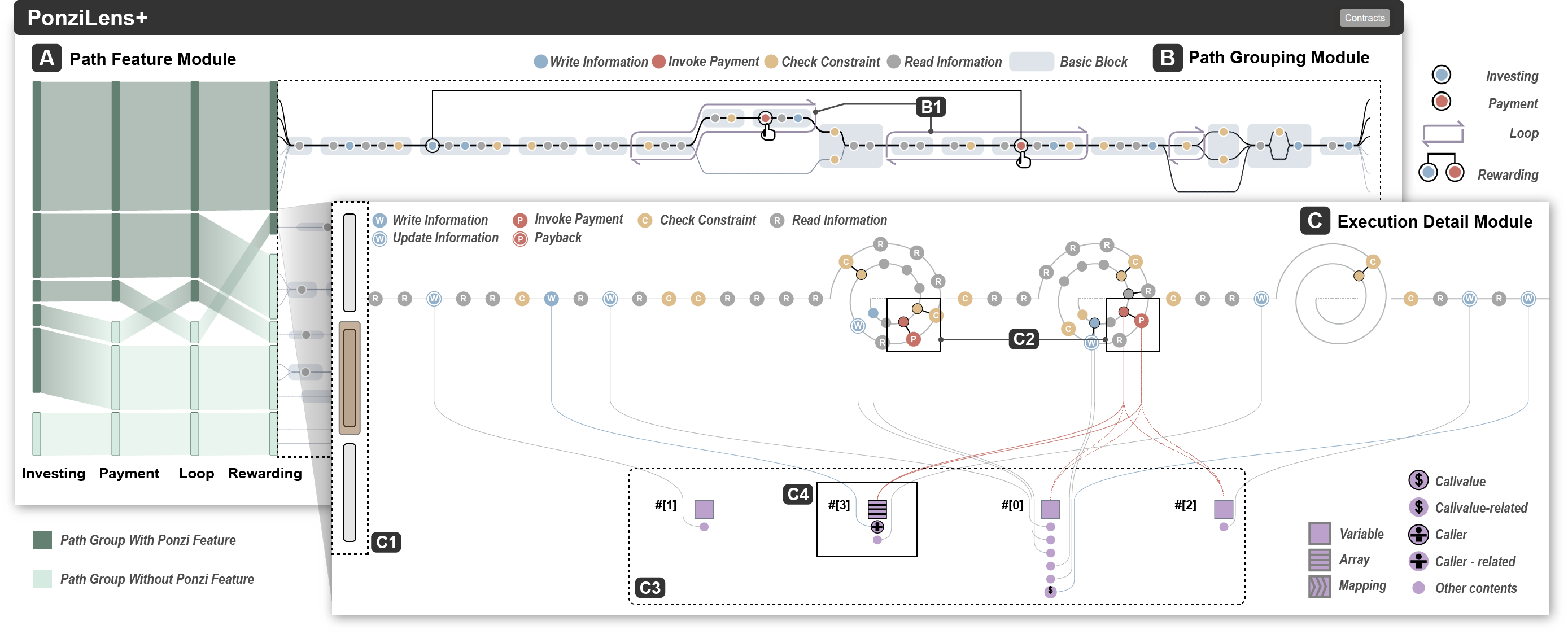}
  \caption{The \techName{} interface initially presents the \Mfeature{} (A) and \Mgroup{} (B).
  Upon selecting specific actions of interest, users can access the \Mpath{} (C), including a scroll bar (C1) to allow users to delve into one single path for more details. }
  \label{design}
\end{figure*}

\section{\techName{}}
Built upon \modify{the semantic action sequences}, we propose \techName{}, a visual analytic system to facilitate the early identification of smart Ponzi schemes.
As shown in Fig.~\ref{workflow}C, \techName{} includes three visualization modules that enable users to diagnose a smart contract at three levels.
\textit{\Mfeature{}} displays the distribution of Ponzi features across all the paths by grouping paths with similar features (R1). 
\textit{\Mgroup{}} offers visual summaries for each path group (R2) and highlights Ponzi features within it (R3). 
\textit{\Mpath{}} presents detailed action sequences and storage interactions during individual executions (R4-R6).
Fig.~\ref{design} shows the user interface of \techName{}, where users initially see the \Mfeature{} and \Mgroup{} and can unfold the \Mpath{} by selecting one or multiple paths from the \Mgroup{}.

\subsection{\Mfeature{}}
The \Mfeature{} (Fig.~\ref{modules}A) offers a quick overview of Ponzi feature distribution across all the execution paths, assisting users in identifying path groups that have a high risk of being a Ponzi scheme.
Initially, the execution paths are grouped based on the four Ponzi features \modify{(\textbf{PF1-PF4}) mentioned in Section~\ref{sec:background} (Fig.~\ref{modules}A1), with paths sharing the same features grouped. 
PF1 and PF4 are marked in S4 of semantic action extraction while PF2 and PF3 are labeled based on the presence of \textit{Invoke Payment} and loops in the action sequences.}
\secondmod{
To visualize the distribution of multiple dimensions (PFs) across various path groups, parallel sets offer a natural visual design since it has been proven effective in visualizing categorical data distribution across multiple dimensions~\cite{kosara2006parallel}.
Parallel sets can also clearly demonstrate the number of paths in each group, which is crucial for users to estimate the proportion of paths with suspicious features.
Moreover, encoding the number of paths in each group through band width helps link to the \Mgroup{}, allowing users to smoothly delve into path behaviors.
Therefore, we adopt parallel sets as the layout for the \Mfeature{}.
Specifically, each column corresponds to a Ponzi feature.
Path groups are represented by bands (Fig.~\ref{modules}A3), with their widths (Fig.~\ref{modules}A2) reflecting the number of paths in each group.
Path groups with a certain feature are indicated in dark green~\darkgreen, while those without are shown in light green~\lightgreen.
These bands traverse four columns and are colored based on whether the groups possess these features.
 }
The total height of the dark (Fig.~\ref{modules}A4) or light green (Fig.~\ref{modules}A5) bars represent the count of paths with or without certain Ponzi features, respectively.

The arrangement of both Ponzi features and path groups is important in the \Mfeature{} since a bad arrangement will make the analysis counter-intuitive or result in band crossings~\cite{kosara2006parallel}.
We ordered the Ponzi features according to the typical flows of manual analysis, starting with ``Investing" and followed by ``Payment" to identify direct redistribution after receiving investments.
``Loop" checking is used next to eliminate irrelevant paths.
``Rewarding", indicative of paying previous investors, is listed as the final feature, as it needs the first two features and helps \modify{determine which group has a high risk for subsequent analysis.
To reduce edge crossing, we first set an initial order by sorting path groups based on whether they contain the four features. Groups with a specific feature are placed above those without it.
}
In practice, users can also interactively adjust the feature order according to their interests by dragging and dropping the Ponzi feature name.

\subsection{\Mgroup{}}
\begin{figure*}[]
  \centering 
    \setlength{\abovecaptionskip}{-0.2cm}
  \setlength{\belowcaptionskip}{-1cm}
  \includegraphics[width=0.95\linewidth
  ]{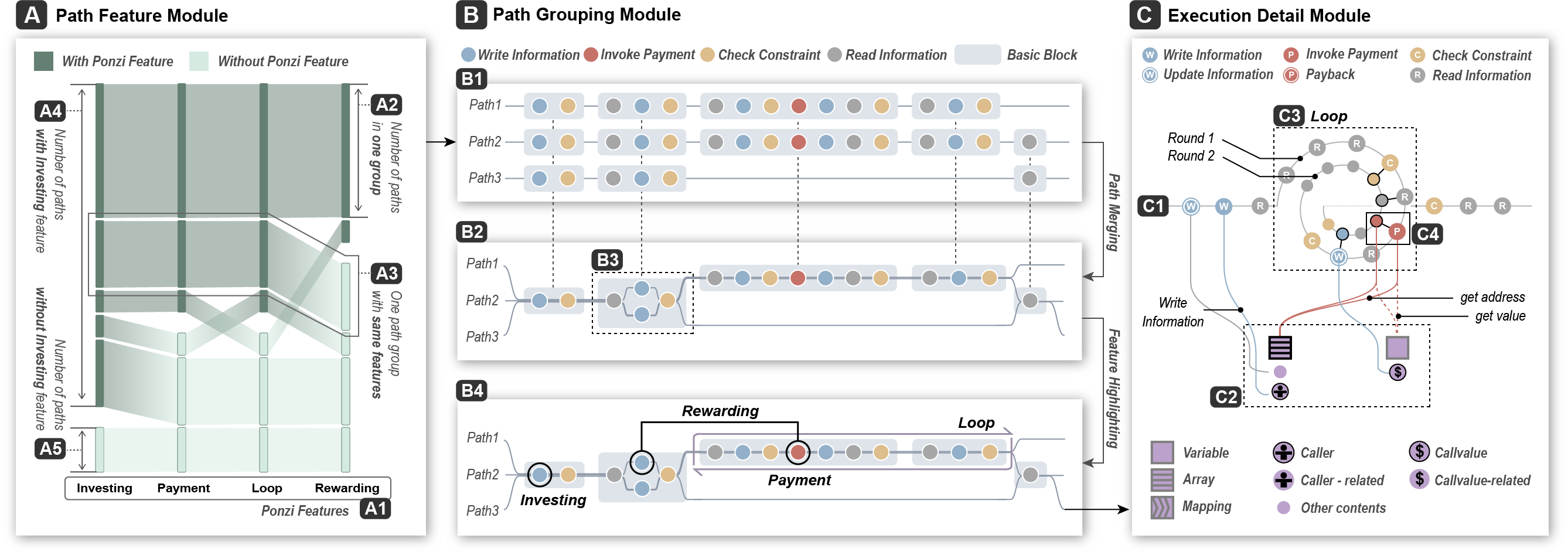}
  \caption{The visual design of three visualization modules in \techName{}. \Mfeature{} (A) shows the Ponzi feature distribution across all execution paths. \Mgroup{} (B) provides a visual summary of action patterns in each path group, incorporating a path merging strategy (B1-B3) and highlighted Ponzi features (B4). \Mpath{} (C) shows the detailed action sequences and storage interactions in each execution path. }
  \label{modules}
\end{figure*}

The \Mgroup{} (Fig.~\ref{design}B) offers visual summaries for each group, where we use circles in four diverse colors to represent four semantic actions (i.e., blue for ``\actW \,\textit{Write Information}", red for ``\actP \,\textit{Invoke Payment}", yellow for ``\actC \,\textit{Check Constraint}", and grey for ``\actR \,\textit{Read Information}"), as shown in Fig.~\ref{modules}B. 
The action sequences are arranged linearly from left to right, encoding the execution order, and are divided according to their basic blocks, with grey blocks~\greyblock \, added behind them for clarity and distinction.

Since displaying all paths is overwhelming, we propose a two-step path-merging strategy \modify{(Appendix A)} to maintain the essential backbone of paths within a single group.
\modify{
First, we group paths that share the same basic block sub-sequence without order conflicts and merge paths based on those basic blocks, as shown in Fig.~\ref{modules}B2. 
Second, we check each merged basic block to verify whether all paths containing this block have the same action sub-sequence, and further separate the paths with different actions within the merged blocks, as shown in Fig.~\ref{modules}B3.
Note that actions of the same type but with different parameters (operands in the stacks during symbolic execution) are also separated. }
\modify{
When drawing this visualization, the x-axis represents the basic block sequence of merged paths, covering all blocks traversed by paths in this group.
The y-axis marks the start and end points of all paths in the group before merging.}
In the middle, the width of each line reflects the number of merged paths.
Therefore, \modify{this module} can provide a comprehensive summary of the action patterns within each group, offering a clear and organized view of the involved execution paths (R2).
Also, we highlight the Ponzi features \modify{(S4 in Section \ref{Sec:action})} in the action summary (R3) to distinctly mark where these features occur during execution, as shown in Fig.~\ref{modules}B4.
The ``\invest Investing" and ``\payment Payment" are highlighted by black circles surrounding blue and red circles, respectively.
``Loop"  is represented by a purple wrapper~\loopicon \, that encases the basic blocks involved in the loop.
To signify the ``rewarding", which means paying previous investors, we connect the ``investing" and ``payment" actions that operate on the same storage slot with a black line \reward.

\subsection{\Mpath{}}

The \Mpath{} (Fig.~\ref{modules}C) displays the detailed action patterns of individual paths (R4). 
Actions are placed from left to right, like those in the \Mgroup{}, as shown in Fig.~\ref{modules}C1.
For clarity, it uses large icons with a letter inside to represent the four actions \actwithletter, which are encoded in the same color scheme as those in the \Mgroup{}.
``\update \textit{Update Information}" and ``\payback \textit{Payback}" \modify{(S4 in Section \ref{Sec:action})} are highlighted with a white border. 

For an in-depth analysis of loops, we collect actions from two rounds of each loop for comparison (R5), since one round is not enough to confirm whether the loop repeats the same actions across different rounds. 
Two rounds are usually adequate for users to spot possible changes, making it unnecessary to include more rounds.
Specifically, the actions within loops in the \Mpath{} are arranged on a two-round Archimedean spiral (Fig.~\ref{modules}C3), with the two rounds placed on the outer and inner circles, respectively.
The Archimedean spiral ensures uniform distance between the same actions of two rounds, facilitating an easy comparison.
The spiral's design aligns with the user's intuitive understanding of loops.
As loops typically traverse the same basic blocks, the two rounds often involve the same sequence of actions, though the same actions may have different parameters.
The actions in the second round (inner circle) are simplified as smaller circles without letters to reduce repeated information.
Those actions with different parameters in the second round are highlighted with black circles and middle lines, as illustrated in Fig.~\ref{modules}C4, which aids users in recognizing the differences quickly.

For visualizing storage interactions \modify{(S3 in Section \ref{Sec:action})}, the storage slots used by the smart contract are depicted beneath the action sequence (as shown in Fig.~\ref{design}C3). 
Three distinct purple square icons represent the typical storage structures: a pure square for variables\,\variable, a square with horizontal lines for arrays\,\arrayicon, and a square with vertical lines for mappings\,\mapping. 
Purple circles\,\content \, are listed under these square icons to symbolize the data stored in these slots, where we use some symbols (as shown at the bottom of Fig.~\ref{modules}C) to highlight the content involving the Ponzi-scheme-related information like the investor's address (\textit{CALLER}~\caller) and amount (\textit{CALLVALUE}~\callvalue).
The contents stored in these slots are connected to the corresponding actions in the action sequence that writes them.
\modify{These connections are represented by lines, colored blue for writing Ponzi-scheme-related information and grey for writing unrelated information.}
In addition, red solid and dashed lines are used to indicate which storage slots the ``\paymenticon \textit{Invoke Payment}" retrieves its parameters from, for the recipient address and payment amount, respectively.
When an action writes \textit{CALLER} in a slot and a payment action retrieves the recipient's address from the same slot, this slot is highlighted in black to denote ``Rewarding" previous investors directly since this slot is used for storing investors' addresses in this smart contract.  
In summary, the \Mpath{} enables users to infer the structure and functionality of each storage slot and to understand the storage interactions of the ``\wicon \textit{Write Information}" and ``\paymenticon \textit{Invoke Payment}" actions.

\begin{figure}[]
  \centering 
    \setlength{\abovecaptionskip}{-0.2cm}
  \includegraphics[width=0.95\linewidth
  ]{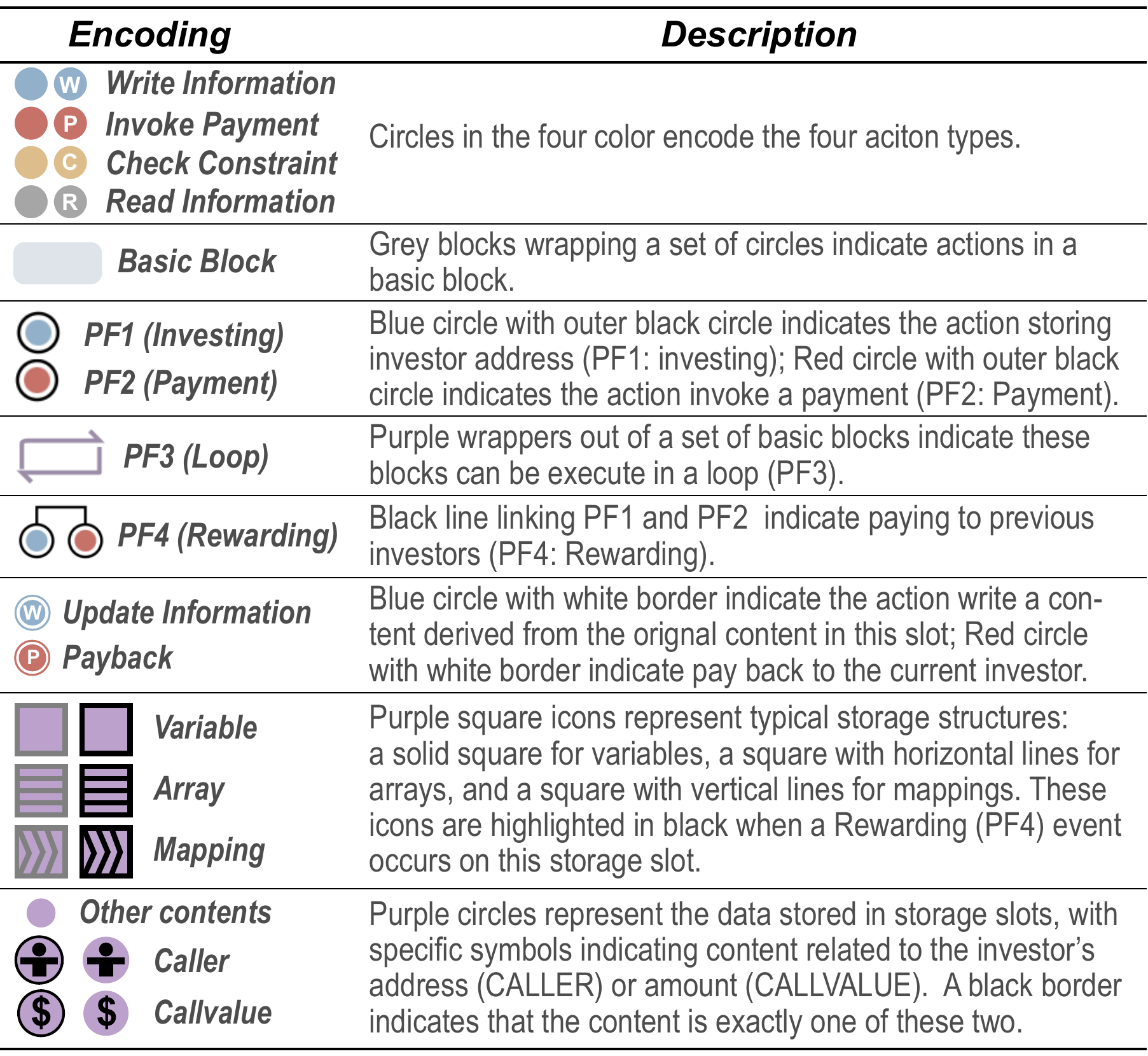}
  \caption{\modify{A summary of visual encoding used in \techName{}.} }
  \label{encodings}
\end{figure}

\subsection{Interactions}
\techName{} enables rich interactions to allow users to \modify{verify whether a smart contract is a Ponzi scheme smoothly.}

\modify{\textbf{Hovering to show details.} 
In \techName{}, users can hover over an action to view a tooltip displaying details of this action collected from S1-S4 in Section~\ref{Sec:action}, such as target slot and content for \textit{Write Information}, payee and value for \textit{Invoke Payment}, and specific constraints for \textit{Check Constraint}.
For storage contents, hovering reveals the underlying Z3 constraints that indicate the content's meaning.}

\modify{\textbf{Navigating to paths of interest.}
\techName{} enables users to easily locate suspicious paths among numerous paths by interacting with three modules. Users can first select a path group with specific Ponzi Features in the \Mfeature{}, then click on actions of interest with specific patterns in the \Mgroup{}. 
These actions are highlighted with a hand icon, and all related paths are marked in black for easy tracking.
Meanwhile, the \Mpath{} appears below to display detailed execution information for the paths containing the selected actions, 
and users can switch between paths using a scroll bar (Fig.\ref{design}C1).
}


\textbf{Comparing two rounds in a loop.}
When users want to check the changes between two rounds of a loop, they are allowed to click the highlighted actions in the inner circle to invoke the tooltip to demonstrate the differences in the parameters between the two rounds, where the same parts are colored in green while the different parts are colored in red.

\modify{The case studies in Section~\ref{sec:cases} will show how to use these interactions in \techName{}.}
\section{Case Study}~\label{sec:cases}
\modify{This section presents two case studies to demonstrate the effectiveness of \techName{} in identifying smart Ponzi schemes, with or without typical features. These case studies were conducted by two users (U4 and U7) of our user interviews, as will be introduced in Section~\ref{sec:interview}.}

\subsection{Case 1: \modify{Scrutinize} a Typical Chain Scheme}
U7 is the creator of a Web3 community and has four years of Web3 investment experience. 
He knows what smart Ponzi schemes are, but lacks experience in auditing the source code.
In the \Mfeature{}, U7 quickly identified a path group possessing all four Ponzi features (Fig.~\ref{design}A), indicating that each path in this group involves investing, rewarding, and at least one loop.
Next, he investigated the visual summary of this group (Fig.~\ref{design}B) and noticed two purple wrappers with a red circle highlighted with a black circle\,\payment, as depicted in Fig.~\ref{design}B1.
It showed that the path group contained two loop-involved payments. 
Further, a red circle in the loop connecting to a blue circle by a black line\,\reward suggests repeated payments to earlier investors, which is a clear sign of a Ponzi scheme.
U7 wondered if the two loops conducted repeated actions, so he clicked to select two payment actions in the loops and unfold the \Mpath{}.
As shown in Fig.~\ref{design}C2, he easily observed that payments in the inner circle were highlighted in both loops, signifying that this path invokes different payments in two rounds of loops.
He also noted that the payments in the first loop had no connections to the storage, whereas those in the second loop retrieved the receiver's address and the payment amount from storage. 
From this, he speculated that the payments of the first loop probably were not to previous investors,
while the payments in the second loop were mainly for distributing money to existing investors.

The chain-scheme is one of the four popular types of smart Ponzi schemes~\cite{chen2021sadponzi}, utilizing a loop to redistribute new investments to each previous investor, whose addresses are stored in a chain-like structure (e.g., an array).
Given that this smart contract involves payments to previous investors in a loop and the slots storing investors' addresses are exactly an array (Fig.~\ref{design}C4), U7 expressed strong confidence in concluding that this contract is indeed a smart Ponzi Scheme, i.e., a chain-scheme.
Such a judgment helped him with an easy decision-making of not investing in this smart contract.

\begin{figure*}[htbp]
  \centering 
    \setlength{\abovecaptionskip}{-0.2cm}
  \setlength{\belowcaptionskip}{-0.4cm}
  \includegraphics[width=0.8\linewidth
  ]{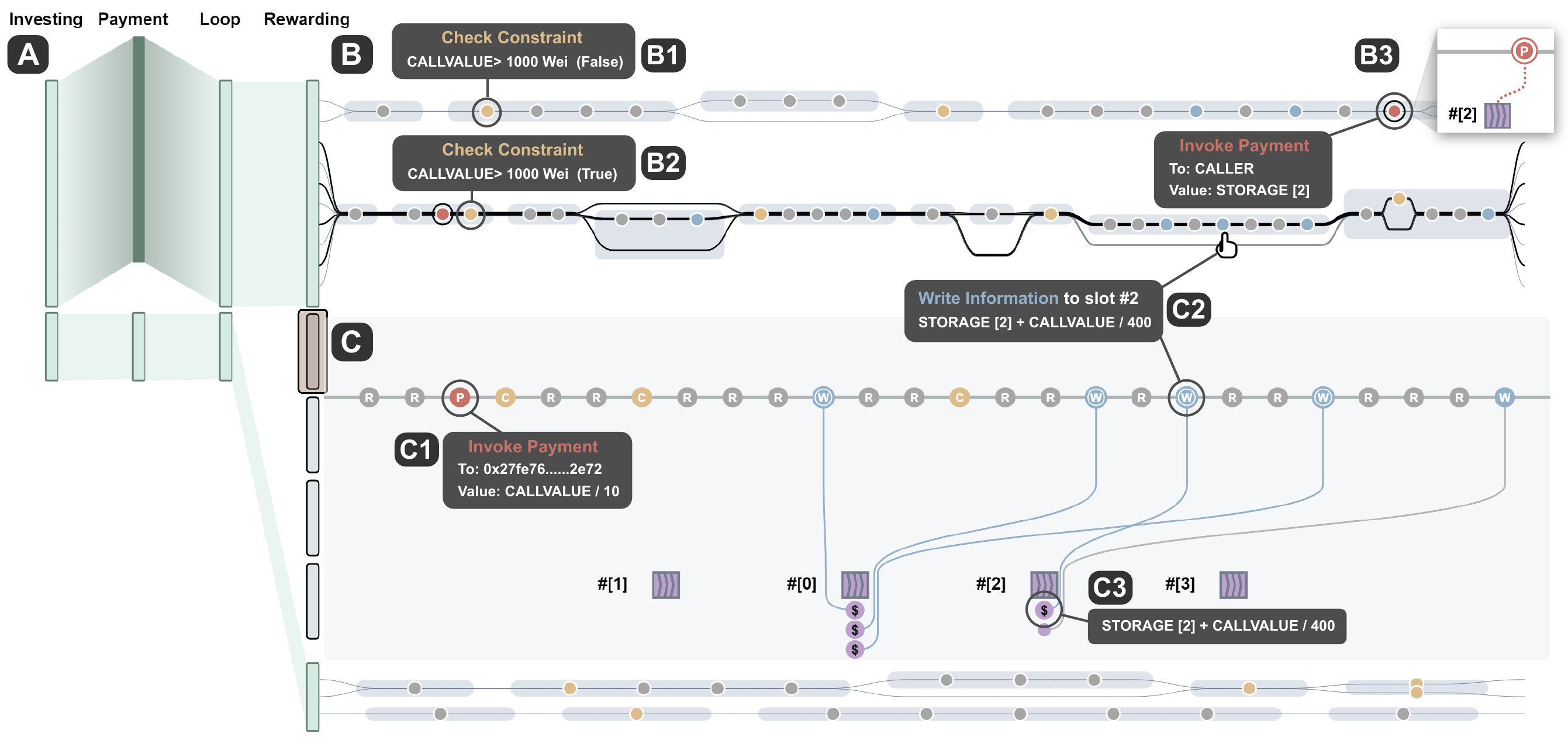}
  \caption{A case for identifying a non-typical smart Ponzi scheme. The \Mfeature{} (A) shows no path groups with multiple Ponzi features. However, the \Mgroup{} (B) shows subgroups of paths with different investing amounts (B1 and B2), and one subgroup pays back the money to the \textit{CALLER} (B3). The \Mpath{} (C) shows more execution details to help conclude this contract is a Ponzi scheme.}
  \label{case3}
\end{figure*}

\subsection{Case 2: \modify{Identify a new variant of smart Ponzi Scheme}}
\label{sec: case 2}

U4 is a smart contract auditor at a Web3 company and one major task of his daily job is to detect vulnerabilities in smart contracts through programs and manual inspection. 
He knows smart Ponzi schemes and the losses they cause to investors, but his work focuses on checking whether smart contracts can be executed successfully instead of frauds like Ponzi schemes.

\modify{Initially,
U4 observed that there were no path groups with multiple Ponzi features and only one path group has the payment feature in the \Mfeature{} (Fig.~\ref{case3}A).}
Thus,
he initially speculated that this smart contract was probably not a smart Ponzi scheme.
But to further validate his judgment, he chose to delve deeper and checked the \Mgroup{} (Fig.~\ref{case3}B). 
He noticed that the paths in the group with payments were divided into two subgroups by our path-merging strategy.
\modify{
By hovering over the two yellow circles (i.e., ``\actC \textit{Check Constraint}''), U4 noticed that both groups had the same constraint (\textit{CALLVALUE$>$1000Wei}) but different conditions, i.e., \textit{False} for the upper one (Fig.~\ref{case3}B1) and \textit{True} for the lower one (Fig.~\ref{case3}B2).
This indicated that if the investment amount was more than 1000 Wei (a measurement unit of the native Ethereum cryptocurrency), the contract would execute paths in the lower subgroup, and vice versa.
After clicking on the payment action in Fig.~\ref{case3}B3 to view the execution details, U4 noticed a red circle enclosed by white\,\payback\, and connected to Slot 2 via a dashed line.
This indicated that the contract returned the money to the current \textit{CALLER}, with the payment amount retrieved from Slot 2.}
U4 confirmed that the paths in this subgroup were probably used to invoke a withdrawal function, returning money to the current \textit{CALLER}.
\modify{
Being curious about how the withdrawn amount was determined, he checked each blue circle to find the ``\actW \textit{Write Information}" that stored the amount in Slot 2 (Fig.~\ref{case3}C2) and clicked it to view the details in the \Mpath{} (Fig.~\ref{case3}C).
The content stored in Slot 2 (Fig.~\ref{case3}C3) showed that the amount that investors can withdraw was increased by \textit{CALLVALUE/400}.
Since these paths were executed only when the investment exceeded 1000 \textit{Wei}, it indicated that the value would increase with each new investment.
Additionally, U4 noticed a ``\paymenticon \textit{Invoke Payment}" (Fig.~\ref{case3}C1) at the start of the path, where \textit{CALLVALUE/10} was paid to a specific address.
He concluded that the smart contract charged ten percent on each investment.}

\modify{The above analysis overturned U4's initial hypothesis, leading him to conclude that the contract was indeed a smart Ponzi scheme, despite lacking typical features like loops or direct rewards.
This contract still exhibited essential Ponzi scheme characteristics, such as maintaining a record of funds owed to prior investors, increasing profit amount based on a portion of new investments, and enabling investors to withdraw profits through additional invocations of the smart contract.}
U4 noted that when the contract balance is sufficient, all investors can receive profits, while the scheme would collapse once the balance becomes limited and unable to sustain payouts to prior investors.
He also emphasized that this kind of contract can easily evade existing rule-based detection methods and that \techName{} was indeed effective in identifying \modify{such a new variant of those typical smart Ponzi schemes.}

\section{User Interview}\label{sec:interview}
We conducted semi-structured user interviews with 12 target users to evaluate the effectiveness and usability of \techName{}.

\subsection{Participants and Apparatus}
We recruited 12 participants (U1-U12) from Web3 communities and universities for our user interviews (2 females, 10 males, $age_{mean}=28.33$, $age_{sd}=4.21$, with normal vision and no color-blindness).
All the participants have enough background in blockchain and smart contracts, and they have experience in investing in smart contracts.
Our participants can be categorized as expert users and common users based on their smart contract auditing experience. 
U1-U6 are experts \modify{skilled in writing and auditing smart contract source code.
U7-U12 are typical investors without such auditing experience, with only U7 being able to understand source code.}
Among them, five participants (U1-U3, U6, U8) are academic researchers specializing in Web3 security, with two (U2-U3) focusing on smart Ponzi scheme detection. 
U4 and U5 have experience in smart contract auditing within Web3 security companies. The other participants (U7, U9-U12) are typical investors with domain knowledge.
\modify{Participants' profiles are shown in Table 1 of Appendix B.}

Our interviews were online via Zoom. We launched the prototype system of \techName{} on the server and allowed participants to assess it via their
own laptops or desktops. Each interview lasted about one hour, and we paid a compensation of $\$$15 to each participant for their time in our user interviews.

\subsection{Procedure }
\begin{figure*}[htbp]
  \centering 
    \setlength{\abovecaptionskip}{-0.1cm}
  \includegraphics[width=0.75\linewidth
  ]{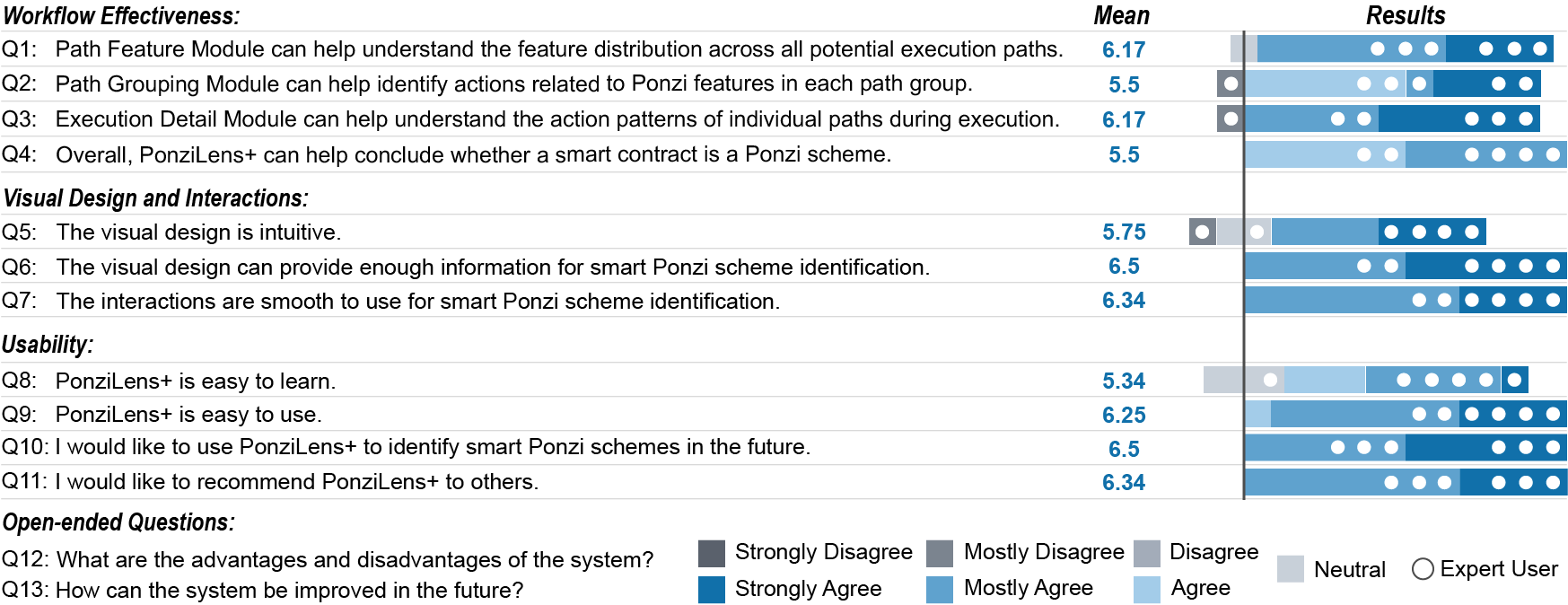}
  
  \caption{The user interview questionnaire results. Q1-Q11 are close-ended questions assessing \techName{}'s workflow effectiveness (Q1-Q4), visual design and interactions (Q5-Q7), and usability (Q8-Q11), rated on a 7-point Likert scale. \modify{Q12-Q13} are open-ended, collecting participant feedback on pros and cons. Results are visualized in a horizontally stacked bar chart, with expert users marked by white circles. }
  \label{results}
\end{figure*}

The interview began with an explanation of the background, visual design, and workflow of \techName{}.
Following this, we presented a usage scenario to the participants, guiding them on how to utilize \techName{} for verifying a smart contract.
The tutorial above lasted about 15 minutes. 
During the interview, the only task was to check whether a smart contract was a Ponzi scheme. 
Hence, we asked participants to leverage \techName{} to verify two smart contracts: one Ponzi contract and one non-Ponzi contract, both randomly selected from a published dataset~\cite{zheng2023securing}.
This task phase had no hard time limit and lasted until participants reached a conclusion, which usually lasted about 30 minutes in practice. 
Finally, we asked participants to finish a post-study questionnaire with 13 questions (Q1-Q13), as shown in Fig.\ref{results}.
Q1-Q11 are close-ended questions that should be answered on a 7-point Likert scale and are designed to evaluate \techName{}'s workflow effectiveness (Q1-Q4), visual design and interactions (Q5-Q7), and usability (Q8-Q11).
Q12-Q13 are open-ended questions to collect participants' feedback on the advantages/disadvantages and possible improvements of \techName{}.
Overall, each user interview session took about 60 minutes. 
All the data collected from the participants were recorded with their permission.

\subsection{Results}
Fig.~\ref{results} presents the participant responses from our post-study questionnaire. 
Overall, \techName{} received high ratings for the close-ended questions. 
Participants unanimously confirmed that there were no existing tools like \techName{} that could offer convincing rationales for identifying smart Ponzi schemes. 
Notably, expert users tended to give higher scores, appreciating \techName{}'s effectiveness in benefiting their contract auditing, as marked in Fig.~\ref{results}.
However, common investors desired more comprehensive tutorials, as they often have a limited background in scrutinizing smart contract execution.
\modify{The main feedback can be summarized as follows:}

\textbf{Workflow Effectiveness.}
Most ratings for Q1-Q4 are positive, which demonstrates that \techName{} has an effective workflow for the visual identification of smart Ponzi schemes. 
Notably, U5 scored ``2" for Q2 and Q3.
He faced challenges in correlating the actions with the smart contract's source code, as his work only involved source code analysis.
U5 acknowledged that symbolically executing bytecodes accurately reflects the actual execution process and can cover all smart contracts on the blockchain. 
However, he pointed out that investors typically lack trust in smart contracts without available source codes, so the source codes can also become suitable inputs to offer extra information for verification.
\modify{Among the three visualization modules (Q1-Q3), the \Mgroup{} received a relatively low mean score (i.e., $5.5/7$).
U8, U9, and U11 found it easy to identify actions marked as Ponzi features, like those in loops or black outer circles, but struggled to assess the usefulness of unhighlighted actions, such as \textit{Check Constraint}, in identifying Ponzi schemes. Therefore, they requested specific examples of unhighlighted actions in tutorials for better guidance.}

\textbf{Visual Design and Interactions.}
All participants agreed that our visual designs are intuitive and provide enough information for identifying Ponzi schemes. Also, they found the interactions user-friendly.
They appreciated the \Mfeature{} and highlighted features \secondmod{in the \Mgroup{}}, which can help quickly understand potential suspicious behaviors.
However, some found \techName{} hard to grasp at the very beginning, particularly in establishing the initial correlation between execution processes and visual encoding.
Once they grasped our semantic action scheme, they acknowledged the design's clarity in showing execution processes.

\textbf{Usability.}
Overall, participants believed that \techName{} was easy to learn and use (Q8-Q9).
After familiarizing themselves with \techName{}, they all expressed willingness to use it in the future and recommend it to others (Q10-Q11).
Both expert users and common investors noted that it took them some time to learn the typical features and action patterns of various Ponzi scheme types due to the lack of knowledge about them.
They suggested that more detailed tutorial documents and examples could be included in \techName{}.

\textbf{\modify{Open-ended Questions}.}
\modify{In response to Q12, all the participants praised \techName{} for offering clear evidence of Ponzi scheme detection, which boosted their analysis confidence.
Additionally, visualizing all potential behaviors greatly aids in understanding the contract's function.
However, they expressed concerns about the learning curve; while investors can easily grasp actions like payments, they may struggle with concepts like storage interaction and stack manipulation. Despite not fully understanding the underlying execution of smart contracts, participants indicated that \techName{} helps them correlate actions with Ponzi features.
For future improvements (Q13), three participants (U2, U6, and U11) proposed adding an initial suspicious score for smart contracts to guide subsequent analysis in \techName{}.}
U3 pointed out that existing Ponzi features and types might not encompass all new Ponzi schemes.
He suggested that \techName{} could be enhanced to allow users to define new features and incorporate them into the \Mfeature{}.
These suggestions are promising and we plan to further improve \techName{} according to them.

\section{Discussion}
In this section, we report the lessons we learned during the development of \techName{}. 
Also, we discuss the generalizability and limitations of \techName{}.

\textbf{Algorithms vs. Visualization.}
Before our work, smart Ponzi scheme detection mainly relied on manual inspection and automatic algorithms~\cite{zheng2023securing}.
The advantage of algorithms is their ability to quickly scan numerous contracts without human effort.
However, they depend on manual labels or predefined rules, which may not cover all Ponzi scheme types and can also result in false alarms.
\secondmod{
During this study, \techName{} has been utilized to check some smart contracts labeled by automatic algorithms in prior research~\cite{chen2021sadponzi} 
and have found some wrong labels provided by the automatic algorithms. One example has been reported in Section~\ref{sec: case 2}). }
Also, we have identified some smart Ponzi schemes that lack typical features or can not be classified into any known Ponzi scheme types.
Further, investors prioritize precise, understandable results for specific smart contracts of interest, rather than checking a large amount of smart contracts at once.
In this context, visualization can play a critical role in revealing the evidence of Ponzi schemes intuitively and additionally aids in both manual inspections and expanding labeled datasets rapidly.

\textbf{Experts vs. Investors.}
\techName{} is designed for both smart contract auditors and common investors hoping to make informed investment decisions. 
Our user interviews revealed different analytical ways between these two types of users.
Common investors particularly appreciate the \Mfeature{} and \Mgroup{}, which enable them to observe the Ponzi feature distribution and check how these features manifest in action sequences by the highlighted Ponzi features.
They usually do not delve into the \Mpath{} to check the storage interactions, except for checking the loop details.
They also recommend adding an initial suspicious score and more examples of action patterns in various Ponzi schemes. 
However, expert users prefer to scrutinize detailed action patterns in both the \Mgroup{} and \Mpath{} to gather more convincing evidence during execution because they find the patterns in the \Mfeature{} a bit general to reach a solid conclusion about whether it is a smart Ponzi scheme.
They also desire to incorporate more information about the source code for mapping action patterns to the actual contract codes.
Overall, \techName{} meets the needs of experts and investors, but it can be improved to further satisfy their customized requirements.

\textbf{Generalizability.}
\secondmod{
Although \techName{} focuses on smart Ponzi schemes, it can be generalized to other applications. 
First, our workflow and visual designs can be easily extended to identify other vulnerabilities and frauds, such as \textit{Multiple Send}~\cite{MS} and \textit{Honeypots}~\cite{torres2019art}, by replacing the Ponzi-specific features with other suspicious behaviors.
Second, our approach of intuitively representing the code execution process as semantic action sequences can assist in broader scenarios involving software code understanding, such as software testing~\cite{zhou2019visfuzz}.
For example, software engineers typically review source code and develop test cases for deeper analysis. \techName{} can expedite this process by providing both an overview and detailed action patterns during software execution. }

\textbf{Limitations.}
As smart contract applications become more extensive and complex, often involving multiple contracts, massive potential execution paths can lead to scalability issues. Our path grouping and merging strategies help mitigate such scalability issues to some extent, and further improvements can be made by grouping paths based on functions and filtering actions by their semantics.
Also, \techName{} targets Ponzi schemes that embed fraudulent logic within smart contract codes, which is common in blockchain-based scams.
However, there are also scams where the fraudulent logic occurs off-chain, such as PlusToken~\cite{zhang2023plus}, which can only be identified through fund flow analysis and is beyond the scope of \techName{}.
\secondmod{
Further, symbolic execution's intrinsic limitations may limit the performance of \techName{}, such as ignoring the gas limits causes exploring paths that are not executable in practice or skipping some paths that the Z3 solver cannot resolve. To mitigate these limitations, we can incorporate gas estimation and employ testing techniques with concrete inputs rather than symbolic ones, such as fuzz testing~\cite{liang2024ponzicuard}.
}



\section{Conclusion}
In this work, we first proposed \modify{a framework to extract} semantically meaningful action sequences to intuitively represent each potential execution path in a smart contract.
\modify{Then, we proposed \techName{}, a visual analytic system for identifying Ponzi schemes based on these semantic actions, which incorporates three visualization modules to demonstrate the action patterns of a smart contract at three different levels and highlight the features related to Ponzi schemes.
We conducted two case studies and in-depth user interviews with 12 users.}
The results demonstrate that \techName{} is useful and effective in assisting users to easily identify smart Ponzi schemes.


In future work, we will enable users to define custom action patterns beyond Ponzi features, expanding \techName{} to broader smart contract testing and auditing.
Furthermore, given that there is an increasing number of cryptocurrency investors using mobile devices (e.g., smartphones) to conduct their investments~\cite{mirza2022mobile}, it is also worth further exploring how \techName{} can be extended to mobile devices.

\section*{Acknowledgments}
This project is supported by the Ministry of Education, Singapore, under its Academic Research Fund Tier 2 (Proposal ID: T2EP20222-0049).


 
\bibliographystyle{IEEEtran}
\bibliography{main}

\vspace{-1cm}
\begin{IEEEbiography}[{\includegraphics[width=1in,height=1.5in,clip,keepaspectratio]{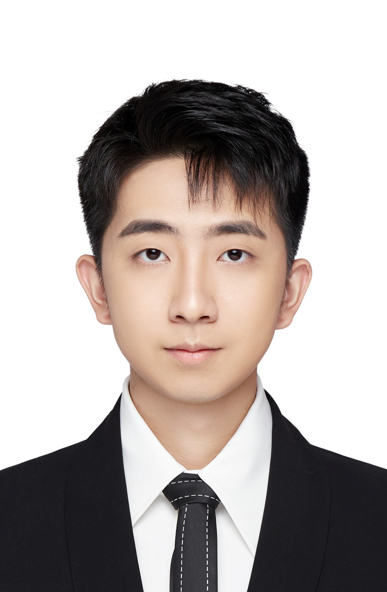}}]{Xiaolin Wen} is currently a Ph.D student in the College of Computing and Data Science at Nanyang Technological University (NTU). 
His research interests mainly focus on visualization and human-computer interaction techniques in Fin-tech and Web3.
He received his master's degree in Computer Science and Technology from Sichuan University in 2023 and his dual bachelor's degree in Computer Science and Financial Engineering from Sichuan University in 2016.
For more information, kindly visit \url{https://wenxiaolin.com/}.
\end{IEEEbiography}

\begin{IEEEbiography}[{\includegraphics[width=1in,height=1.5in,clip,keepaspectratio]{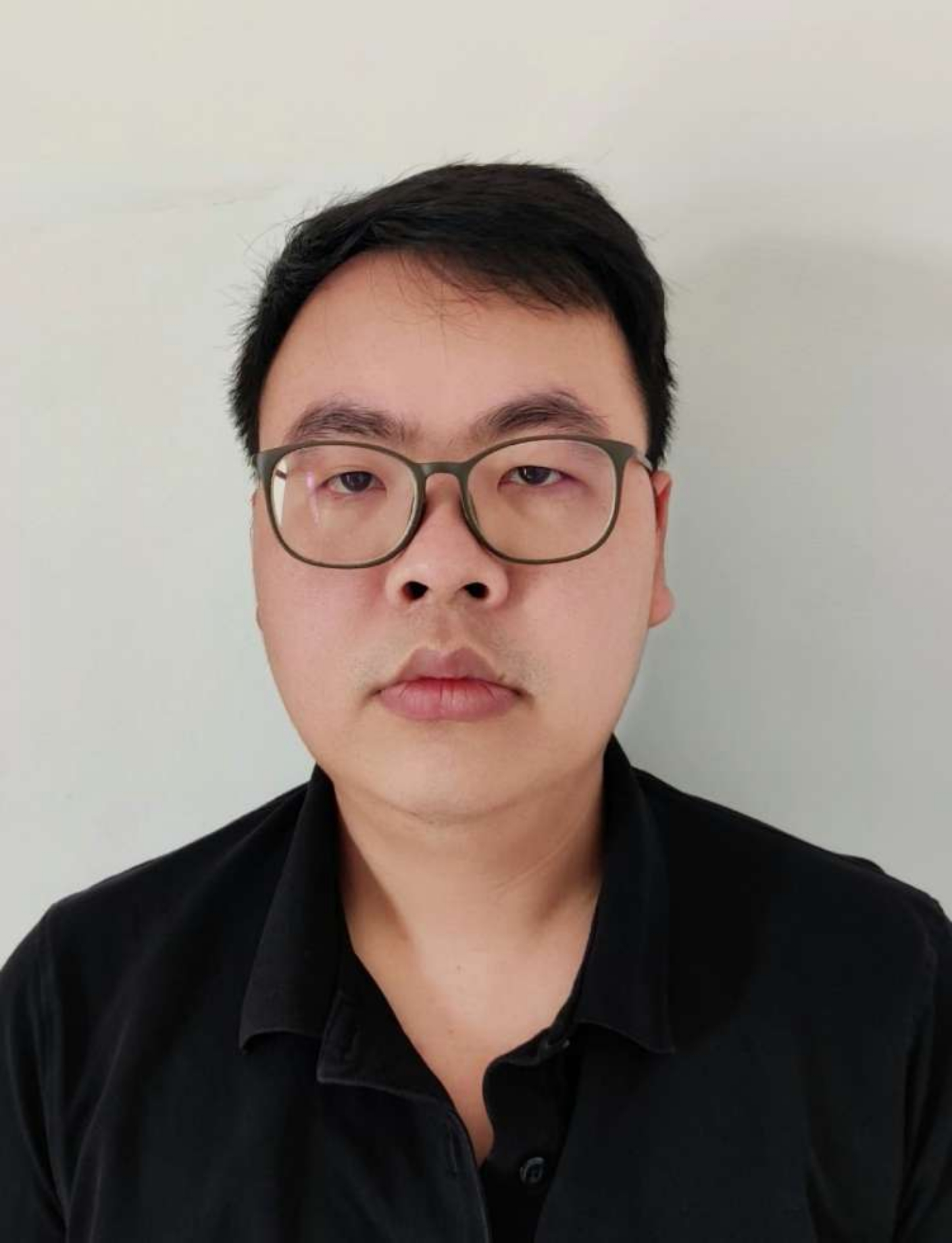}}]{Tai D. Nguyen} received his Ph.D. degree in computer science from the School of Computing and Information Systems at Singapore Management University (SMU). He is currently a research fellow at SMU. His current research interests include formal methods, smart contract security, and AI security. For more information, kindly visit \url{https://duytai.github.io/}.
\end{IEEEbiography}

\begin{IEEEbiography}[{\includegraphics[width=1in,height=1.5in,clip,keepaspectratio]{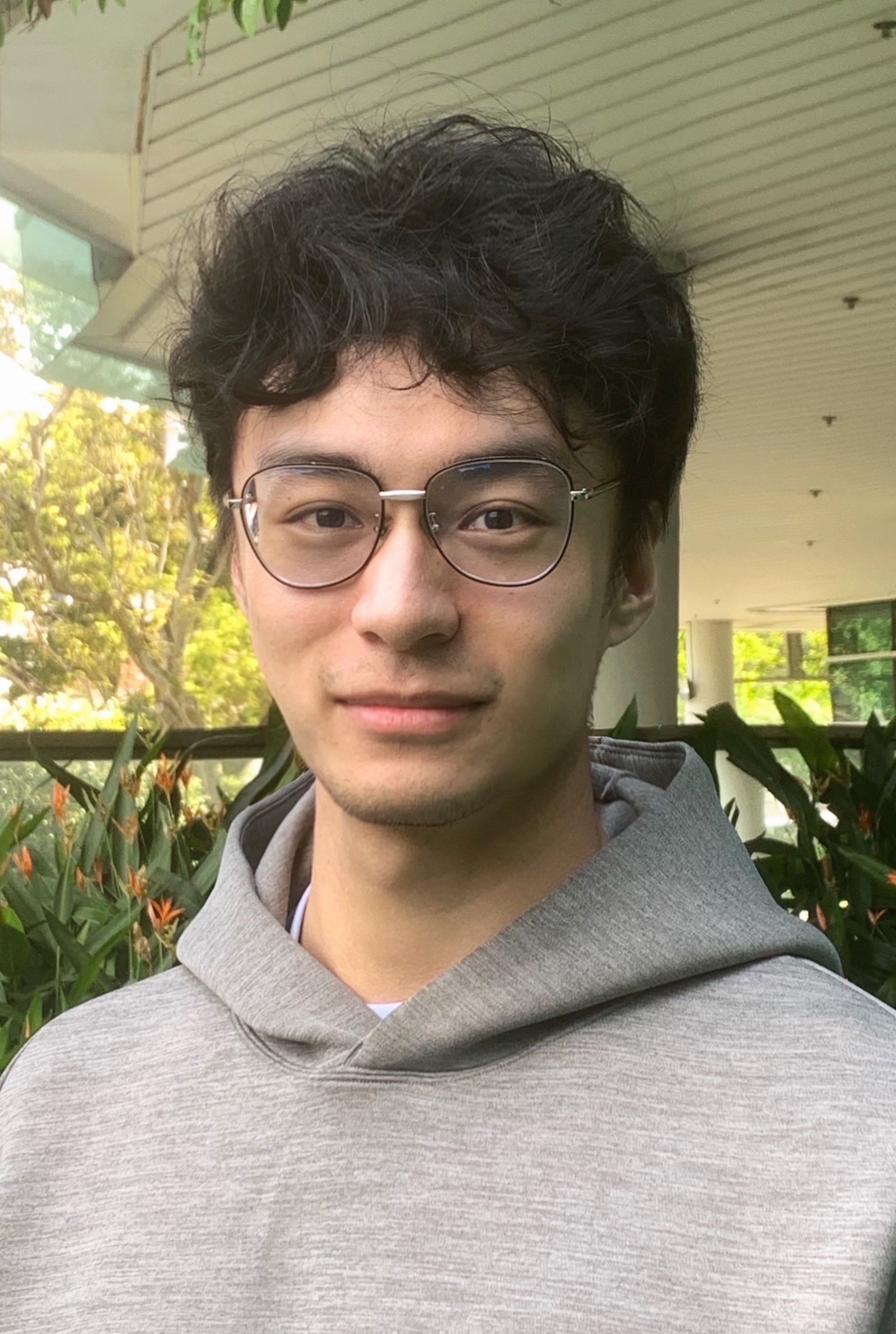}}]{Shaolun Ruan} is currently a Ph.D. candidate in School of Computing and Information Systems at Singapore Management University (SMU). 
His work focuses on developing novel graphical representations that enable a more effective and smoother analysis for humans using machines, leveraging the methods from Data Visualization and Human-computer Interaction.
He received his bachelor's degree from the University of Electronic Science and Technology of China (UESTC) in 2019.
For more information, kindly visit \url{https://shaolun-ruan.com/}.
\end{IEEEbiography}

\begin{IEEEbiography}[{\includegraphics[width=1in,height=1.5in,clip,keepaspectratio]{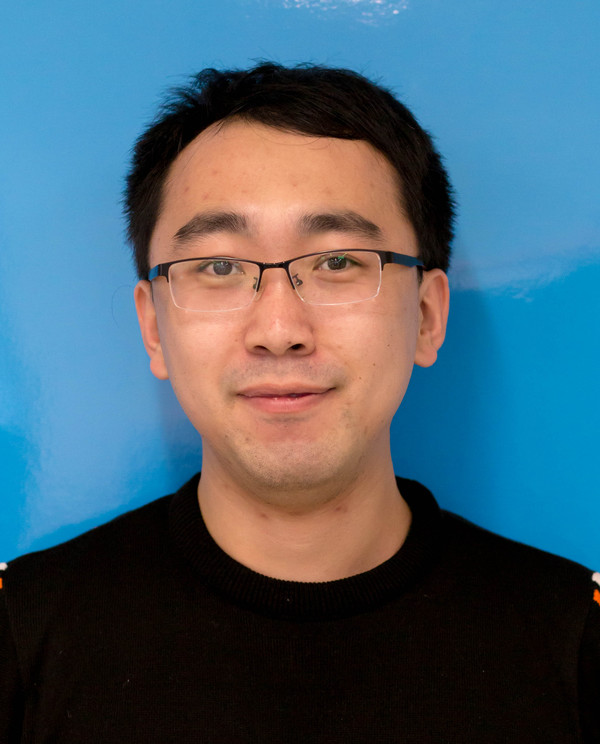}}]{Qiaomu Shen} received the PhD degree in computer science from The Hong Kong University of Science and Technology in 2020. He is currently an research assistant professor in Southern University of Science and Technology(SUSTECH).  Before joining SUSTECH, he worked as a senior developer at Noah's Ark Lab for Huawei Technologies.  His current research interests include spatial-temporal visualization, urban computing, and visual analytics of complex system.
\end{IEEEbiography}

\begin{IEEEbiography}[{\includegraphics[width=1in,height=1.5in,clip,keepaspectratio]{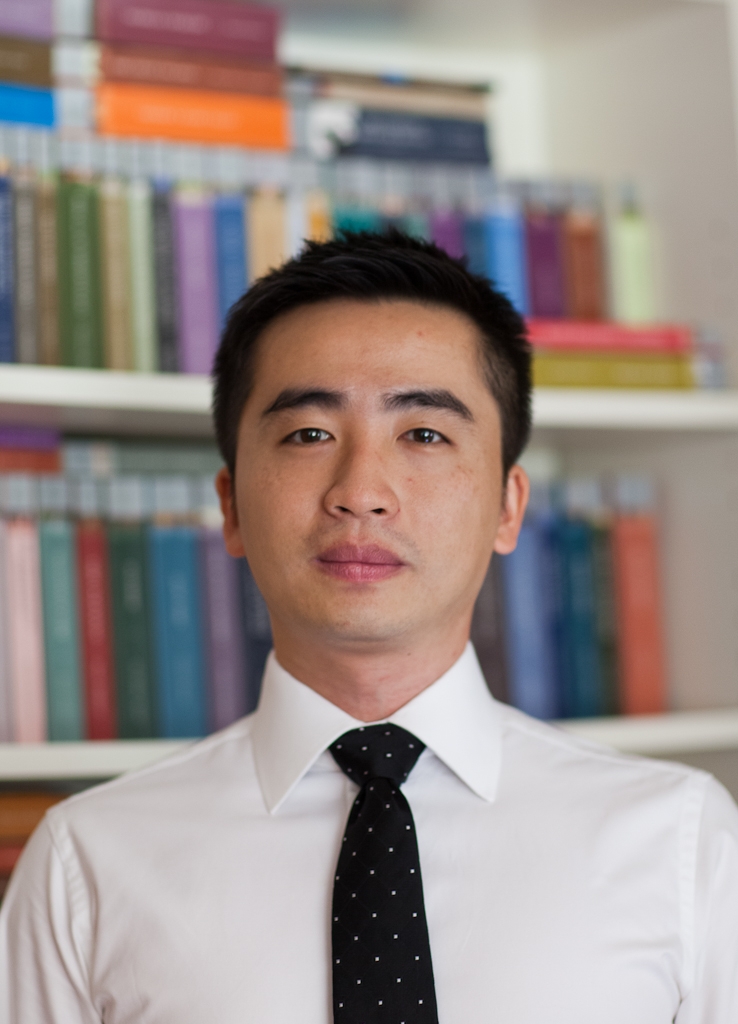}}] {Feida Zhu} is currently a tenured Associate Professor
and Associate Dean at the School of Computing
and Information Systems, Singapore Management
University. 
He obtained his Ph.D. in Computer Science from
the University of Illinois at Urbana-Champaign (UIUC) in 2009.
His research interests include AI and
collaborative intelligence, blockchain, data asset and
AI governance, with emphasis on their application
to business, financial and consumer innovation. Prof.
ZHU has over 100 peer-reviewed research publications
at top international venues, including ICDE,
VLDB, SIGMOD, KDD, WWW, JMLR, TODS,
TKDE, etc. 

%
\end{IEEEbiography}

\begin{IEEEbiography}[{\includegraphics[width=1in,height=1.5in,clip,keepaspectratio]{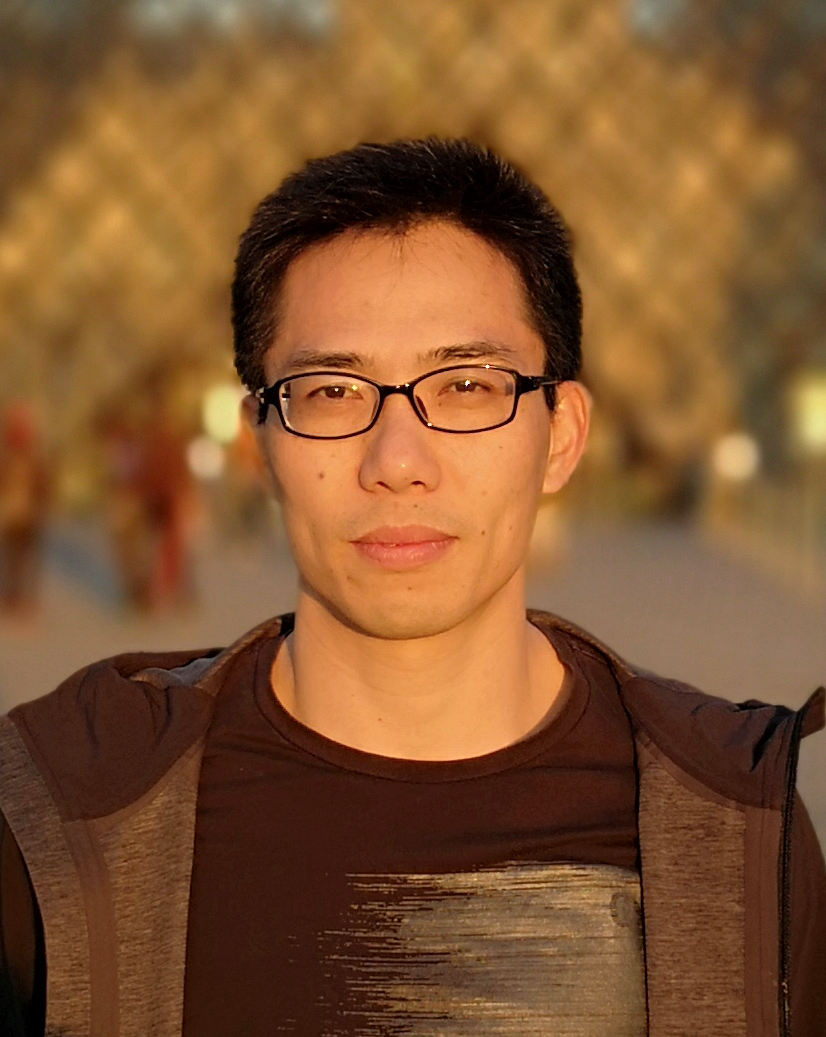}}]{Jun Sun}is currently a professor at Singapore Management University (SMU). He received Bachelor and PhD degrees in computing science from National University of Singapore (NUS) in 2002 and 2006. He has been a faculty member since 2010. He has received the prestigious LEE KUAN YEW fellowship twice in 2007 and 2022. Jun's research interests include formal methods, program analysis and lately AI security. His profile can be found at \url{https://sunjun.site}.
\end{IEEEbiography}

\begin{IEEEbiography}[{\includegraphics[width=1in,height=1.5in,clip,keepaspectratio]{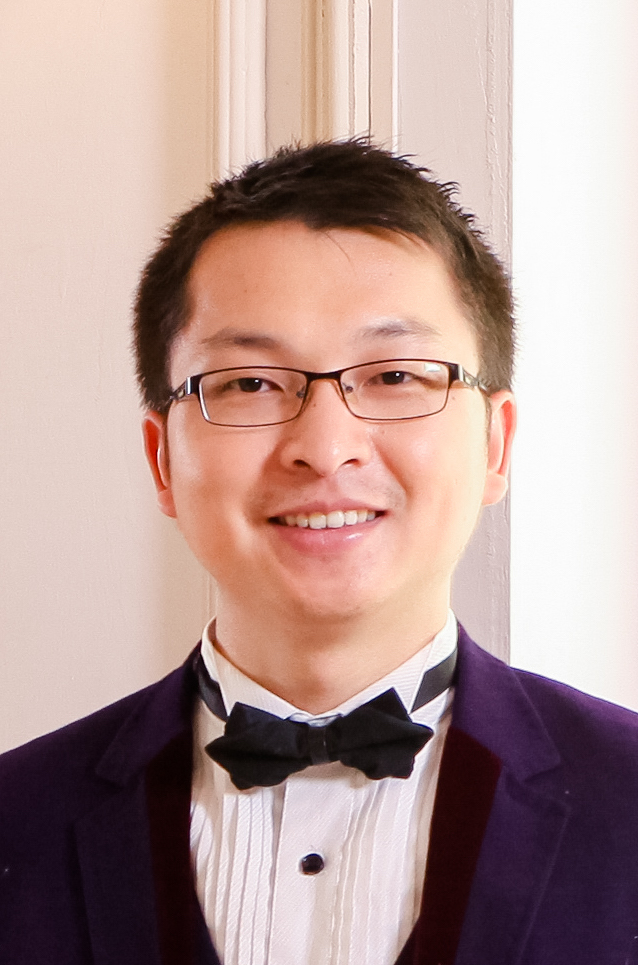}}]{Yong Wang} is currently an assistant professor in the College of Computing and Data Science, Nanyang Technological University. Before that, he worked as an assistant professor at Singapore Management University from 2020 to 2024. His research interests include data visualization, HCI and human-AI collaboration, with an emphasis on their application to FinTech, quantum computing and online learning. He obtained his Ph.D. in Computer Science from Hong Kong University of Science and Technology. He received his B.E. and M.E. from Harbin Institute of Technology and Huazhong University of Science and Technology, respectively. For more details, please refer to \url{http://yong-wang.org}.
\end{IEEEbiography}


\newpage
\onecolumn
\appendix
\section*{A. Path Merging Strategy}

In the appendix, we describe the details of our \textbf{Path Merging Strategy} used in \Mgroup, as shown in Algorithm~\ref{Appendix: Merging}.
Our path-merging strategy has two steps: 
\textbf{Step 1: Merge paths sharing the same basic block sub-sequence.} 
Represent each execution path by its basic block number sequence, group paths with identical sub-sequences and no order conflicts, and merge them to create the longest sequence that includes all basic blocks of these paths.
\textbf{Step 2: Separate paths with different actions in merged basic blocks}. 
Compare the semantic action sequences of different paths in the merged basic blocks from Step 1. 
If different actions are found within the same basic block, separate them to maintain individual actions.
After the above two steps, we visualize the merged execution paths in the Path Group Module. 

\SetKwComment{Comment}{/* }{ */}
\begin{algorithm}[hbt!]
\label{Appendix: Merging}
\SetAlgoLined  
\caption{Path Merging Strategy}\label{alg:two}
\KwIn{$L$: Execution paths list, where each path consists of a sequence of basic blocks, and each basic block contains a sequence of actions.}
\KwOut{$MergedPaths$: List of merged paths, where the same basic blocks are merged and the different actions in merged blocks are separated.}
\Comment{\textbf{Step1: Merge paths sharing the same basic block sub-sequence}}
\Comment{Map each path in $L$ into a sequence of basic block index}
$BasicBlockSequences \gets MapToBasicBlockSequences (L)$\;
\Comment{Divide basic block sequences into sub-groups by the rule: sharing the same basic block sub-sequence without order conflicts}
$BasicBlockSequencesSubGroups \gets GroupByRule (BasicBlockSequences)$\; 

$FullSequenceList \gets $ []\Comment*[r]{List of full sequence for each sub-group}
$MergedPaths \gets $ []\Comment*[r]{List of merged paths with different actions separated}
\Comment{For each sub-group of paths that can be merged}
\For{each $subGroup$ in $BasicBlockSequencesSubGroups$}{
\Comment{Merge all basic block sequences of a sub-group into the longest full sequence}
$FullSequence \gets MergeIntoFullSequence(subGroup)$\;

$FullSequenceList \gets Add(FullSequence)$\;
\Comment{\textbf{Step2: Separate paths with different actions in merged basic blocks}}
$Differences \gets $ []\Comment*[r]{Different actions in the merged paths}
\For{each $BasicBlock$ in $FullSequence$}{
\Comment{Get a set of paths containing this basic block from the sub-group}
$PathsSet \gets GetPathByBlock(BasicBlock,subGroup)$\;
\Comment{Fetch action sequences of this basic block from the original path data}
$ActionSequences \gets GetActionSequences(PathsSet,L)$\;

$Difference \gets CompareActions(ActionSequences)$\;
\Comment{Compare the differences in actions among paths }
$Differences \gets Add(Difference)$
}
$MergedPaths \gets Add(FullSequence, Differences)$
}
\textbf{Return} $MergedPaths$
\end{algorithm}

\newpage
\section*{B. Participant Information Table}
In the user interviews, we gathered participant profiles, including gender, age, and experience with web3 applications and smart contract auditing. We also asked them to describe how smart contracts relate to their daily work. 
All the participants have enough background in blockchain and smart contracts, and they have experience in investing in smart contracts. 
For participants who preferred not to describe their specific jobs, we just marked them as ``A web3 investor who has invested in smart contracts" in this table.

\begin{table*}[hbt!]
\caption{\label{participant}
The detailed information of the user interview participants.
All participants are experienced in smart contract investment. 
U1-U6 have experience in auditing smart contracts, while U7-U12 lack experience in smart contracts auditing.
}
\resizebox{\textwidth}{25mm}{
\begin{tabular}{cccccl}
\hline
ID & Gender & Age & Web3 Experience & Auditing Experience & Description\\\hline

    U1    &    Male    &   31  &   60 months    &  48 months &  A PhD expertise in smart contract security, working at a web3 security company.   \\
    U2    &     Male   &    27 &    36 months  & 24 months & A PhD candidate with a published paper on smart Ponzi scheme detection.    \\
    U3    &    Male    &  29   &  42 months & 6 months &  A PhD candidate in blockchain analysis, interning at a web3 security company.     \\
    U4    &    Male    &  35   &   30 months  & 24 months & A smart contract auditor at a web3 security company.      \\
    U5    &    Male    &   34  &  18 months    &  8 months  &  A team leader of web3 projects at an internet company.      \\
    U6    &     Male   &  23   &    6 months   & 6 months &  A master student working on smart Ponzi scheme detection.  \\

    \hline

    U7   &      Male  &   32  &      48 months   & 0 months   &  A creator of a web3 community and a key opinion leader on Twitter.  \\
    U8    &   Male     &   25  &     36 months  & 0 months &    A PhD candidate expertise in Cryptography.   \\
    U9    &    Male    &  30   &  24 months  & 0 months & A web3 investor who has invested in smart contracts.     \\
    U10    &    Female    &   27  &    18 months    &  0 months & A web3 investor who has invested in smart contracts.  \\
    U11   &     Female   &  22   &    12 months   & 0 months &  An operations manager at a web3 security company.    \\

    U12   &     Male   &   25  &   10 months     &  0 months &  A web3 investor who has invested in smart contracts.    \\

    \hline
\end{tabular}
}
\end{table*}

\end{document}